\documentclass[pre,amsmath,amssymb,aps,reprint,groupedaddress,nofootinbib,raggedbottom,floatfix]{revtex4-2}
\pdfoutput=1
\usepackage[T1]{fontenc}
\usepackage[utf8]{inputenc}
\usepackage{lmodern}
\usepackage{scalerel}
\usepackage{gensymb}

\usepackage[citecolor=blue, colorlinks=true, pdfusetitle, pdfauthor={Surajit Chakraborty}, pdfsubject={Statistical Physics of Disordered systems}, pdfkeywords={Amorphous solids, heterogeneous elasticity, long-range correlation, jamming}]{hyperref}

\begin{document}

\title{Long-Range Correlations in Elastic Moduli\\ and Local Stresses at the Unjamming Transition}

\author{Surajit Chakraborty} 
\email{schakraborty@tifrh.res.in}

\author{Kabir Ramola} 
\email{kramola@tifrh.res.in}

\affiliation{Tata Institute of Fundamental Research, Hyderabad 500046, India}

\date{\today}

\begin{abstract}
We explore the behaviour of spatially heterogeneous elastic moduli as well as the correlations between local moduli in model solids with short-range repulsive potentials. We show through numerical simulations that local elastic moduli exhibit long-range correlations, similar to correlations in the local stresses. Specifically, the correlations in local shear moduli exhibit anisotropic behavior at large lengthscales characterized by pinch-point singularities in Fourier space, displaying a structural pattern akin to shear stress correlations. Focussing on two-dimensional jammed solids approaching the unjamming transition, we show that stress correlations exhibit universal properties, characterized by a quadratic $p^2$ dependence of the correlations as the pressure $p$ approaches zero, independent of the details of the model. In contrast, the modulus correlations exhibit a power-law dependence with different exponents depending on the specific interaction potential. Furthermore, we illustrate that while affine responses lack long-range correlations, the total modulus, which encompasses non-affine behavior, exhibits long-range correlations.

\end{abstract}

\maketitle

\section{Introduction}
Solids differ from fluids in their ability to resist deformation in response to external forces~\cite{landau2013fluid}. Amorphous solids, such as glasses and granular solids, lack long-range order similar to fluids, yet they are rigid in response to small mechanical perturbations~\cite{cates1998jamming,schuh2003atomistic}. Unlike crystalline materials where the elasticity is derived from a symmetry-broken, unique stress-free state,  amorphous solids lack such a reference state. The rigidity of these materials emerges from the complex, heterogeneous contact networks that form between particles, satisfying conditions of local mechanical equilibrium. Due to the inherent disorder within the material, amorphous solids show large spatial fluctuations in their mechanical properties, characterized by localized stresses and heterogeneous local elastic moduli~\cite{alexander1998amorphous}.

Despite their varying microscopic details that differ among different amorphous structures and result in different physical properties, amorphous solids exhibit several common mechanical properties. Stress fluctuations in amorphous solids display long-range anisotropic correlations, decaying as a power-law of $1/r^d$ in $d$ dimensions \cite{lois2009stress,lemaitre2017inherent,lemaitre2018stress,degiuli2018field,henkes2009statistical,lerner2020simple}. These correlations are anisotropic at large lengths scales exhibiting pinch point singularities in Fourier space~\cite{nampoothiri2020emergent,nampoothiri2022tensor,vinutha2023stress}. The observed rigidity of amorphous solids has also been linked to the existence of such long-range stress correlations originating from random reference states \cite{yoshino2010emergence,tong2020emergent,nampoothiri2022tensor}. Amorphous solids are also known to exhibit regions termed ``soft spots'', where the stiffness associated with collective vibrations is small. These low-frequency vibrations are considered (quasi-) localized as only a limited number of particles near the core participate in the vibration~\cite{lerner2021low}. The presence of these localized vibrations leads to non-affine displacements of particles within the material when subjected to deformation about mechanical equilibrium. These non-affine displacements make a negative contribution to the local moduli~\cite{lemaitre2006sum, hentschel2011athermal,cui2019theory}. The combined effect of non-affine displacements and inherent disorder gives rise to local stiffness variations  when subjected to deformations, resulting in a spatially heterogeneous elastic response~\cite{mizuno2013measuring, mizuno2016elastic, mizuno2016spatial,pogna2019tracking,wagner2011local,fan2014evolution,leonforte2006inhomogeneous,yoshimoto2004mechanical,shakerpoor2020stability,caroli2019fluctuating,gelin2016anomalous}. In this context, understanding the nature of elastic heterogeneity is crucial for understanding the anomalous elastic properties exhibited by disordered solids. 

Systems composed of soft particles, which are largely unaffected by temperature fluctuations, enter a jammed state when the density or applied stress reaches a threshold value \cite{liu1998jamming}. This jamming transition signifies the onset of rigidity within a disordered assembly, and is exhibited across various structures like foams, emulsions, granular solids, and glasses~\cite{liu1998jamming, trappe2001jamming,gopal2003relaxing,chaudhuri2007universal,bi2011jamming,ikeda2012unified,katgert2013jamming,charbonneau2014fractal}. Theoretical frameworks for jammed amorphous solids extensively usually rely on configurations of frictionless particles with finite-range repulsive interactions at zero temperature~\cite{makse2000packing,o2002random,silbert2002statistics,o2003jamming,wyart2005rigidity,wyart2005geometric,silbert2005vibrations,henkes2007entropy,henkes2009statistical,ellenbroek2009jammed,wyart2012marginal,goodrich2016scaling,ramola2017scaling}. Athermal arrangements of soft, repulsive frictionless spheres or disks demonstrate a well-defined jamming transition by varying the packing density at zero applied shear stress. 
At zero pressure, the system achieves isostaticity, marked by an average number coordination of $2d$ per particle, where $d$ represents the dimensions. Below this threshold, the system loses its rigidity. Approaching the transition, floppy modes dominate the linear response~\cite{wyart2005geometric,silbert2005vibrations}, and the system exhibits diverging length scales and scaling behavior evident in both mechanical as well as geometrical properties~\cite{o2002random,o2003jamming,ellenbroek2006critical}. Such materials exhibit anomalous elastic properties as the unjamming transition is approached, with the bulk modulus scaling as $p^{(\alpha - 2)/(\alpha - 1)}$, while the shear modulus scales as $p^{(\alpha - \frac{3}{2})/(\alpha - 1)}$~\cite{o2002random,o2003jamming,van2009jamming}. Here, $\alpha$ represents the exponent associated with the repulsive interaction, with the potential energy scaling as $v(\delta)\sim \delta^{\alpha}$, where $\delta$ denotes the interparticle overlap. Although global moduli have been extensively investigated, the local elastic moduli and its heterogeneity within such materials has received relatively less attention.

\begin{figure}[t!]
\centering
  \includegraphics[width=1\columnwidth]{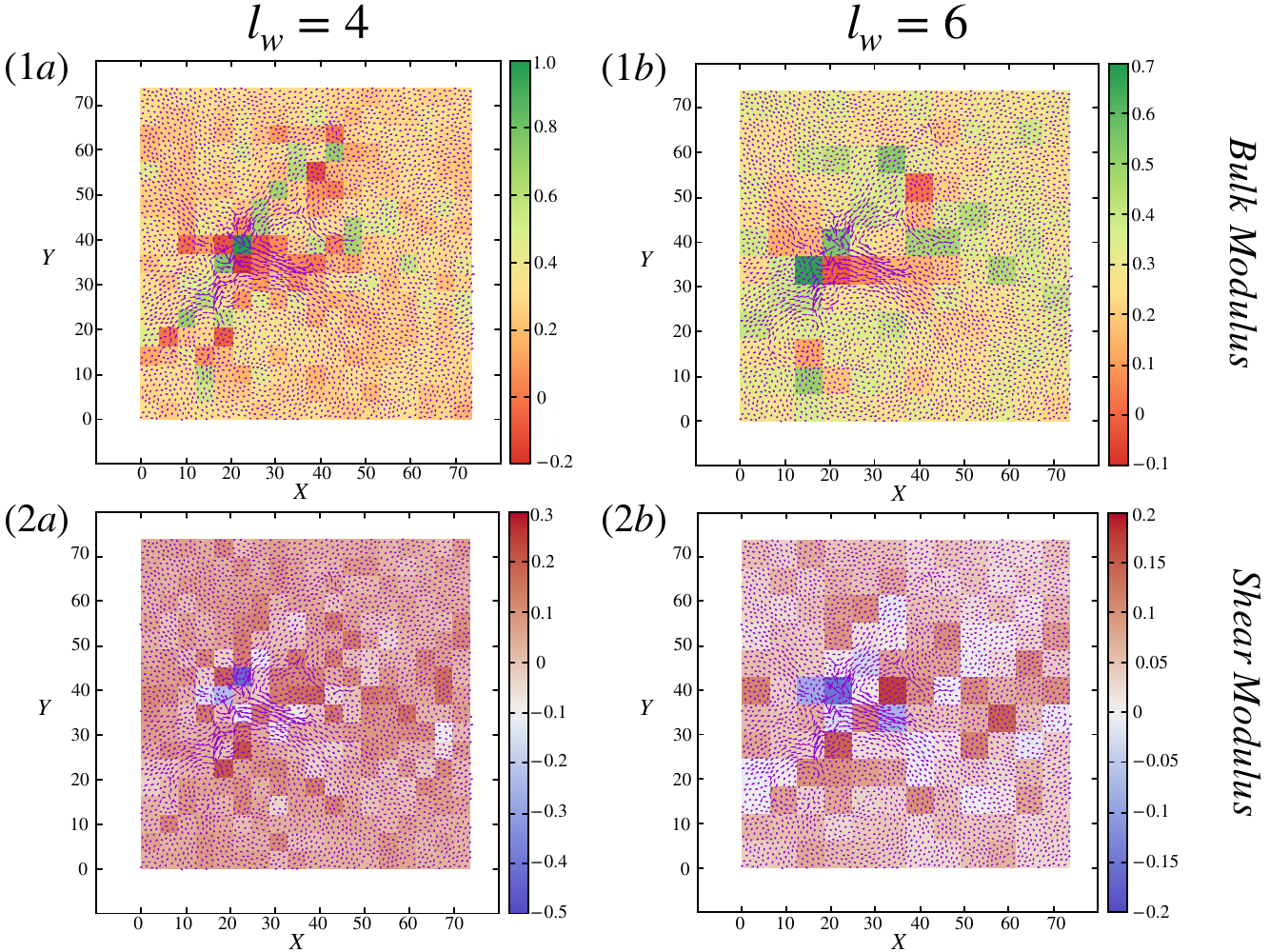}
  \caption{ (Color map) Spatial map of the local elastic modulus ((\textbf{1}) bulk modulus ($K_m$) and (\textbf{2}) shear modulus ($G_m$)) for a particular configuration of $N=4096$ harmonically interacting disks at a pressure of $10^{-2}$. The map is presented for different coarse-grained box sizes, specifically (\textbf{a}) $l_w=4$ and (\textbf{b}) $l_w=6$. The vector field overlaid on the map illustrates the first non-zero vibrational mode of the system. Local bulk and shear modulus show large fluctuations near regions with significant particle displacements in low-energy excitations. Details regarding the measurements are provided in the text.
  }
  \label{elastic_moduli_heterogeneity_mode}
\end{figure}

The anomalous mechanical, thermal, and acoustic properties observed in amorphous solids \cite{zeller1971thermal,pohl2002low,buchenau1984neutron,ruffle2003observation,masciovecchio2006evidence,monaco2009anomalous} have been suggested to originate from the presence of spatially heterogeneous elastic moduli~\cite {schirmacher2006thermal,schirmacher2007acoustic,schirmacher2015theory,marruzzo2013heterogeneous,mizuno2013elastic,mizuno2014acoustic,mizuno2016relation,gelin2016anomalous}. This local heterogeneity is illustrated in Fig.~\ref{elastic_moduli_heterogeneity_mode} where we show the spatial map of (\textbf{1}) local bulk modulus and (\textbf{2}) local shear modulus for a particular configuration of $N=4096$ harmonically interacting disks at a pressure of $10^{-2}$. The map is presented for different coarse-graining box sizes ($l_w$) utilized in defining the local moduli (details are provided in the subsequent text), specifically (\textbf{a}) $l_w=4$ and (\textbf{b}) $l_w=6$. The vector field overlaid on the map illustrates the first non-zero vibrational mode of the system. Local bulk and shear modulus show large fluctuation caused by large non-affine contribution to the corresponding modulus.

Some recent studies on three-dimensional jammed solids and low-temperature glasses have suggested that local elastic moduli exhibit negligible correlations at large lengthscales~\cite{mizuno2016spatial,mizuno2018phonon,shakerpoor2020stability}. In contrast, other studies investigating acoustics properties in low-temperature glasses such as Gelin~\textit{et al.}~\cite{gelin2016anomalous} suggest long-range spatial correlations in elastic moduli. Theoretical frameworks often invoke the presence of extended power-law correlations in elastic moduli~\cite{cui2020analytical,cui2020vibrational}. However, Mizuno \textit{et al.}~\cite{mizuno2018phonon}, utilizing a protocol similar to that in~\cite{gelin2016anomalous} for measuring affine elastic moduli, reported the absence of long-range spatial correlations in elastic moduli. In contrast, Mahajan \textit{et al.}~\cite{mahajan2021emergence} investigated the local moduli in three dimensional disordered solids by measuring the local stress response to global strain deformation, and observed anisotropic correlations with a $r^{-3}$ decay in the local shear moduli measured in this manner. In a more recent study, Zhang \textit{et al.} investigated local elastic moduli in jammed solids and observed long-range behavior above a threshold pressure~\cite{zhang2023local}. Thus the nature of the correlations in local elastic moduli in amorphous systems remains a subject of some debate.

In this study, we show the presence of long-ranged correlations in local elastic moduli. Our analysis focuses on the spatially heterogeneous elastic moduli of two-dimensional amorphous solids composed of soft disks that interact via short-range repulsive interactions. To explore the correlations between elastic moduli, we measure the correlations in Fourier space and observe that the fluctuations in local elastic moduli exhibit long-range correlations, similar to the stress tensor correlations that are anisotropic and decay as $1/r^2$ at large lengthscales. 
Specifically, the local (simple) shear modulus correlations reveal anisotropic behavior with pinch-point singularities in Fourier space, resembling the structural patterns observed in shear stress correlations. We characterize both the affine and non-affine contributions to the modulus correlations in Fourier space, specifically focusing on small wavenumbers.

In addition, we investigate the fluctuations in local stresses and moduli near the unjamming transition across model solids with short-range repulsive potentials. Our results reveal that stress fluctuations exhibit a universal characteristic, with correlation magnitude decays as $p^2$ when the pressure $p$ approaches the unjamming point, independent of specific interaction details. In contrast, modulus correlations exhibit model-dependent behavior, following power-law decay with exponents associated with the specific interaction potential. Fluctuations in the local affine moduli scale as its spatial average, i.e. $\langle \lvert \delta K_m \rvert \rangle \sim \langle \bar{K}_m \rangle$, as pressure $p$ decreases towards the unjamming point. In contrast, the total shear modulus, which includes significant non-affine contributions, exhibits anomalous scaling behavior. The relative fluctuations in local shear modulus ($\langle \lvert \delta G_m \rvert \rangle / \langle \bar{G}_m \rangle$) distributions increase following distinct power-law exponents as the system approaches the unjamming transition, highlighting the critical influence of non-affine deformations on the mechanical properties of amorphous solids.

The outline of the paper is as follows. The first section describes the numerical protocol employed to generate two-dimensional amorphous packings at fixed global pressures. In the second section, we show the correlation of the local stress tensor as the unjamming transition is approached. The measurement of local elastic moduli heterogeneity within the two-dimensional solids is presented in the third section. In the fourth section, we investigate the correlations of the (simple) shear moduli in Fourier space. Finally, we conclude by summarizing our observations and highlighting the significance of our results.

\begin{figure*}
\centering
\includegraphics[width=2\columnwidth]{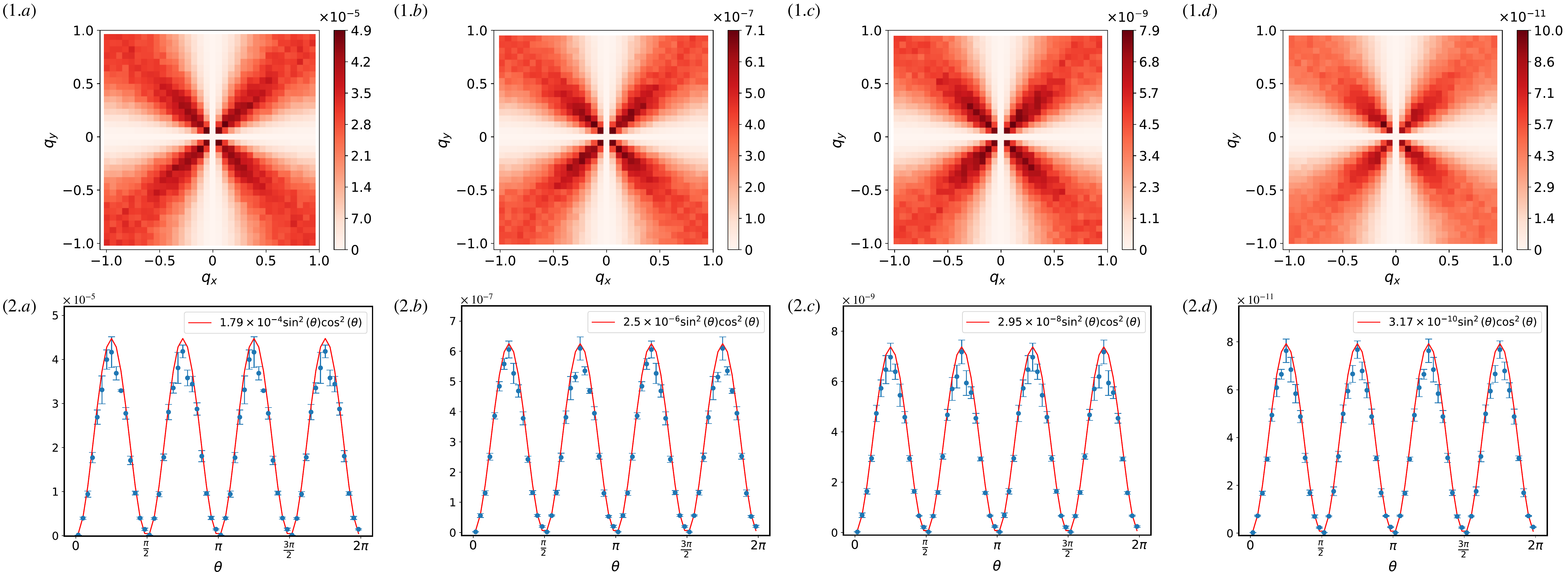}
\caption{Shear stress correlations within two-dimensional amorphous packings under Hertzian repulsion are studied in Fourier space at varying pressures: (\textbf{a}) $p=10^{-3}$, (\textbf{b}) $p=10^{-4}$, (\textbf{c}) $p=10^{-5}$, and (\textbf{d}) $p=10^{-6}$. The stress tensor in real space is derived using~\eqref{eqn_stress_def} with coarse-grained boxes of size $l_w = 3$. In the top row, we present a two-dimensional representation of shear stress correlations. These reveal pinch-point singularities as $\lvert q \rvert \rightarrow 0$ at all pressures, indicating the presence of long-range anisotropic correlations extending up to the unjamming point. The bottom row illustrates the angular dependence of the correlations at small $\lvert q \rvert$, computed through radial averaging within $\lvert q \rvert = 2\frac{2 \pi}{L}$ to $10\frac{2 \pi}{L}$ using angular intervals of $\frac{2\pi}{60}$. These correlations exhibit an angular dependency of $A \sin^2\theta \cos^2\theta$ for various $A$ values at different pressures.}
\label{stress_correlation_figure}
\end{figure*}

\section{Model and Methods}\label{sec_model}
We simulate jammed packings of a bi-disperse ($50:50$) mixture of frictionless disks with a diameter ($a$) ratio of $1:1.4$ in two dimensions with periodic boundary conditions. The particles interact via a short-range repulsive interaction of the form
\begin{equation}  
v(r_{ij})=\frac{\epsilon}{\alpha}\left(1-\frac{r_{ij}}{a_{ij}}\right)^\alpha,
\end{equation} 
where $r_{ij}$ is the center-to-center distance between disk $i$ and disk $j$, whereas $a_{ij}=\frac{a_i + a_j}{2}$ is the sum of the radii of the two disks. The small particle diameter ($a_0$) is the unit of length and $\epsilon$ is the unit of energy. The coarse-grained stress components and moduli are expressed in units of $\frac{\epsilon}{a_0^3}$. We study canonical models of jamming with frictionless soft particles, such as the Harmonic ($\alpha=2$) and Hertzian ($\alpha=\frac{5}{2}$) models. Additionally, we explore potentials related to different $\alpha$ values. We utilize the Conjugate Gradient minimization protocol to generate energy minimized arrangements at different pressures following a variant of the standard O'Hern protocol~\cite{o2002random,goodrich2012finite}. The ensembles are characterized by the pressure of the configurations. Unless explicitly mentioned, all analyses are conducted on solid configurations with a system size $N=8192$ grains. Grains identified as ``rattlers'', characterized by a lack of overlap with neighboring grains, are subsequently removed from the analysis.

\section{Stress Correlations at the unjamming transition}\label{sec_stress_correlation}

Granular solids exhibit long-range anisotropic stress correlations, indicating the presence of force chains within these materials~\cite{nampoothiri2020emergent,nampoothiri2022tensor}. Long-range anisotropic stress correlations have also been observed in low-temperature glasses, attributed to the percolation of force-bearing networks~\cite{tong2020emergent}. Moreover, the existence of long-range anisotropic stress correlation has been identified in rigid gels~\cite{vinutha2023stress}, revealing identifiable force chains within these materials. In this study, we investigate stress correlations in two-dimensional jammed solids as they approach the unjamming transition, examining the decay of these correlations across model solids with short-range repulsive potentials. \\
In order to investigate the local mechanical properties, specifically the stress tensor and elastic modulus tensor, we follow previously established coarse graining procedures~\cite{mizuno2013elastic,mizuno2016spatial,shakerpoor2020stability}. We partition the simulation box into coarse-graining boxes of different sizes, each characterized by a linear length $l_w$ measured in units of the smallest particle diameter. The stress tensor is obtained by coarse-graining the force moment tensor. The force moment tensor associated with the contact between particles $i$ and $j$ is defined as the cross product of the contact force vector $\vec{f}_{ij}$ and the distance vector $\vec{r}_{ij}$. The local stress tensor within a coarse-grained box $m$ is determined by considering the fraction of bond lengths contained within that box, defined as,
\begin{equation}           
\sigma_{\alpha\beta}^m=\frac{1}{l_w^2} \sum_{i<j} \frac{\partial v(r_{ij})}{\partial r_{ij}} \frac{r^\alpha_{ij} r^\beta_{ij}}{r_{i j}} \frac{q_{ij}}{r_{ij}} \label{eqn_stress_def}.
\end{equation}
Here, $\frac{q_{ij}}{r_{ij}}$ represents the fraction of the bond length $r_{ij}$ contained within box $m$.

To extract the long-range behavior of stress correlations at large length scales, we measure the correlations in the Fourier domain. Typically, such long-range correlations are easier to analyze in the Fourier domain, since one only has to analyze the behavior in a small range near $\lvert q \rvert \to 0$, as opposed to a large region in real space which can also produce a large amount of noise in the data (since the values are small at large distances). We measure stress correlations in Fourier space by performing a discrete Fourier transform of the fluctuations in the local stress tensor component $\Delta\sigma_{\alpha\beta}^m$, representing the difference between $\sigma_{\alpha\beta}^m$ and its spatial average, $\bar{\sigma}_{\alpha\beta}^m$:
\begin{equation}
 \Delta\sigma_{\alpha\beta}(\vec{q})= \sum_{m}\exp(i \vec{q}.\vec{r}_m ) \Delta\sigma_{\alpha\beta}^m.
\end{equation}
Here, $\vec{q}$ represents the wavevector in Fourier space and $\vec{r}_m$ denotes the coordinates of the center of the $m$-th box. The correlation at $\vec q$ is given by,
 \begin{equation}
 C_{\alpha \beta \gamma \delta}(\vec q ) = \langle\Delta\sigma_{\alpha \beta}(\vec q )\Delta\sigma_{\gamma \delta}(-\vec q )\rangle,
 \end{equation}
where `$\langle\rangle$' denotes average over configurations. 
Each point in Fourier space is separated by $\frac{2\pi}{L}$, where $L$ represents the length of the square simulation box, and the upper cutoff is determined by the size of the coarse-grained box. To measure the correlation, we average over a minimum of 200 configurations for each pressure.

\begin{figure}[]
\centering
  \includegraphics[width=1\columnwidth]{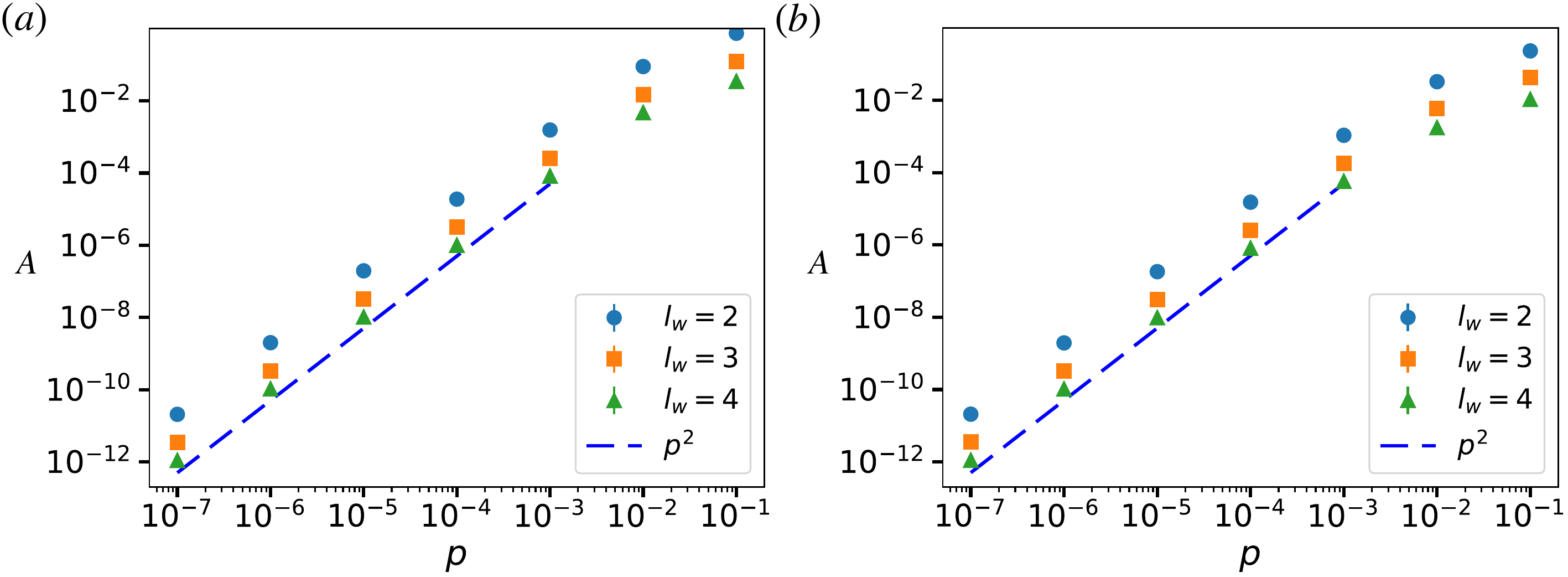}
  \caption{Dependence of the amplitude ($A$) of shear stress correlations, which behave as $C_{xyxy}(q,\theta) = A \sin^2\theta \cos^2\theta$, on the pressure ($p$) of the packings. Remarkably, the stress correlation exhibits a universal behavior across different model solids. In both models, $A$ follows a quadratic dependence on pressure ($p^2$) at small pressures as shown in \textbf{(a)} and \textbf{(b)} for Harmonic ($\alpha=2$) and Hertzian ($\alpha=\frac{5}{2}$) respectively.}
  \label{stress_correlation_strength}
\end{figure}

In Fig.~\ref{stress_correlation_figure}, we present the Fourier space representation of shear stress correlations across different pressures. The top row of Fig.~\ref{stress_correlation_figure} illustrates the two-dimensional representation of these correlations. The correlation is anisotropic at large length scales and exhibits different limits in Fourier space as $|q| \to 0$ is approached from different directions. This feature is known as a ``pinch point'' singularity. Such ``pinch point'' singularities have been observed in the context of spin correlations within frustrated magnetic systems~\cite{prem2018pinch}. Pinch-point behavior has also been identified as a significant feature of stress correlations in granular systems~\cite{henkes2009statistical,mcnamara2016eshelby,degiuli2018field,wang2020connecting}. Recent theoretical studies predict that pinch-point singularities in stress correlations of amorphous solids arise from mechanical equilibrium conditions~\cite{nampoothiri2020emergent,nampoothiri2022tensor}. Notably, stress correlations unveil pinch-point singularity at $\lvert q \rvert \rightarrow 0$ across all pressures, signifying the presence of long-range anisotropic stress correlations extending up to the unjamming point. Such anisotropic structures, characterized by pinch-point singularities as $\lvert \mathbf{q} \rvert \rightarrow 0$ within stress correlators, have been observed in both simulations and experimental studies of jammed solids~\cite{nampoothiri2020emergent}. A recently developed theoretical framework, based on the mapping between the elasticity theory of granular solids and tensorial electromagnetism provides the functional expression for the observed anisotropic stress correlations~\cite{nampoothiri2020emergent,nampoothiri2022tensor},
\begin{equation}
C_{xyxy}(q,\theta) = A \sin^2\theta \cos^2\theta.
\label{k_2d_eqn}
\end{equation}
The angular dependence of shear stress correlations is shown in the bottom row of Fig.~\ref{stress_correlation_figure}. The angular dependence of the correlations at small $\lvert q \rvert$ are computed through radial averaging within $\lvert q \rvert = 2\frac{2 \pi}{L}$ to $10\frac{2 \pi}{L}$ using angular intervals of $\frac{2\pi}{60}$.
Next, we investigate the amplitude of the anisotropic stress correlations ($A$) in model solids with short-range repulsive potentials as they approach the unjamming transition.

In Fig.~\ref{stress_correlation_strength}, we illustrate the behaviour of $A$ as the system approaches the unjamming point, employing different coarse-graining box sizes for stress measurement in real space. Notably, increasing the coarse-graining box length diminishes the magnitude of fluctuations, impacting the overall correlation magnitude. Interestingly, we observe a universal quadratic scaling ($p^2$) of stress correlation amplitude ($A$) regardless of the specific microscopic details. This dependence on the pressure can be explained from phenomenological arguments: as the local stress tensor is the first moment of the force distribution, at low pressures all elements are proportional to $p$, yielding a $p^2$ dependence for the stress correlations, independent of the form of interactions. This simple analysis is also corroborated by a more detailed field theory constructed based on the symmetries observed in jammed solids under isotropic compression predicts, at the transition point, the stress correlation vanishes as the square of the pressure ($p^2$) of the packings~\cite{henkes2009statistical}. Recent studies on near-crystalline packings have provided a more microscopic derivation of this scaling, and shown that it is robust across an amorphization transtion~\cite{maharana2024universal}. This behaviour has also been observed in systems with finite ranged Lennard-Jones interaction~\cite{PhysRevE.96.032902}. 

\begin{figure}[]
\centering
  \includegraphics[width=1\columnwidth]{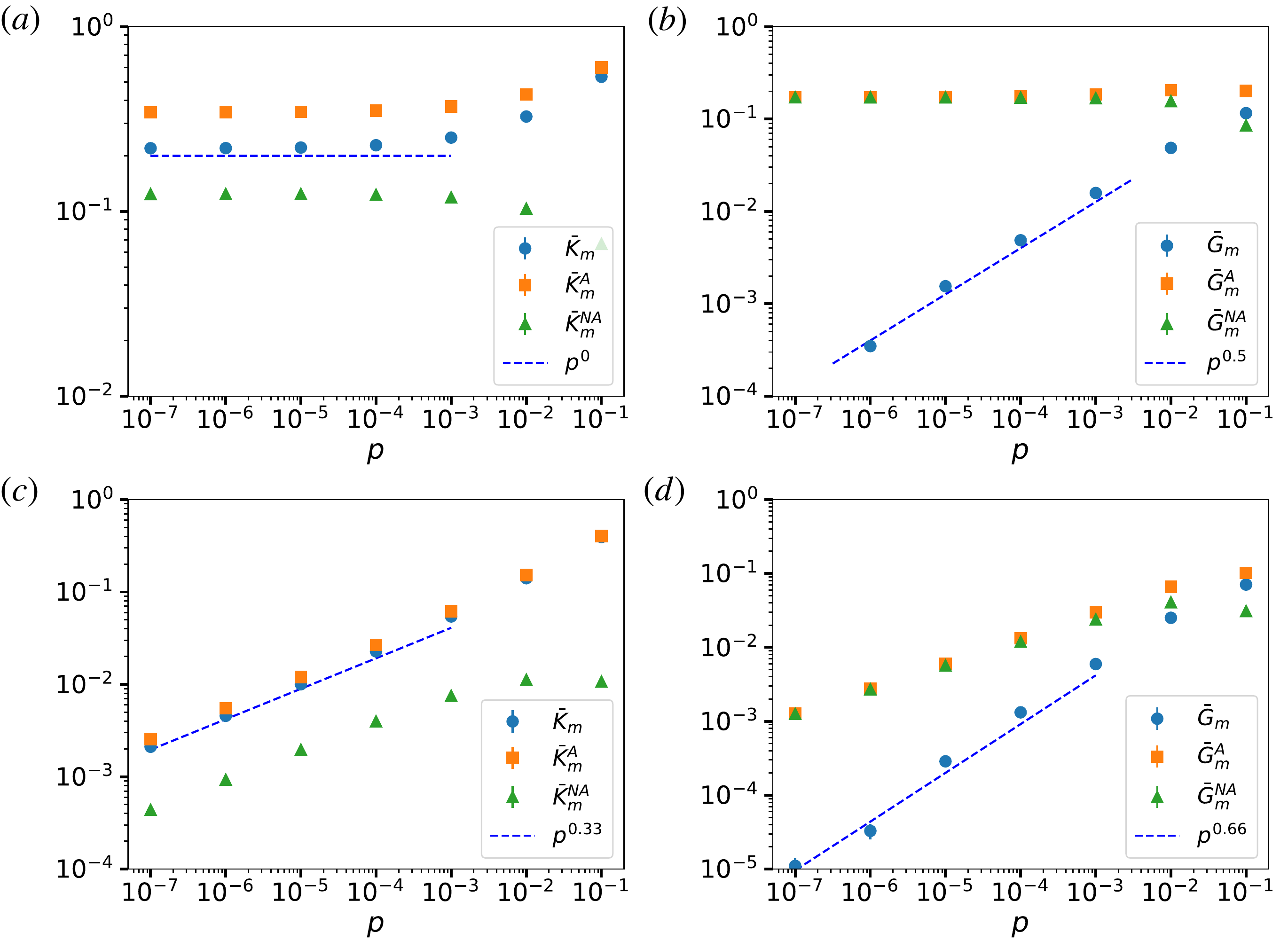}
  \caption{The pressure-dependent behaviors of mean bulk modulus ($\bar{K}_m$) and mean shear modulus ($\bar{G}_m$) are shown for different particle interaction potentials (open circles). For Harmonic interactions ($\alpha=2$), the mean bulk modulus ($\bar{K}_m$) remains constant, while the mean shear modulus ($\bar{G}_m$) scales as $p^{1/2}$ at low pressures, as shown in \textbf{(a)} and \textbf{(b)} respectively. In the case of Hertzian interactions ($\alpha=5/2$), the mean bulk modulus ($\bar{K}_m$) scales as $p^{1/3}$ and the mean shear modulus ($\bar{G}_m$) scales as $p^{2/3}$, as illustrated in \textbf{(c)} and \textbf{(d)} respectively. The corresponding behaviors for affine ($\bar{G}^A_m$, $\bar{K}^A_m$) and nonaffine ($\bar{G}^{NA}_m$, $\bar{K}^{NA}_m$) moduli are represented by open triangles and squares respectively. The local moduli are evaluated using a coarse graining length of $l_w = 3$.
}\label{elastic_modulus_mean}
\end{figure}
\section{Heterogeneous Elastic Moduli} \label{elastic_moduli_hetrogeneity}

Among various alternative derivations to compute elastic moduli~\cite{PhysRevE.107.064608,gelin2016anomalous,mizuno2016spatial}, in this study, we measure elastic responses by assuming hypothetical affine deformations of the solid. We measure the local elastic moduli using the ``fully local'' approach described in~\cite{mizuno2013elastic,mizuno2016spatial} under athermal conditions ($T=0$). This evaluation incorporates both affine (A) and non-affine (NA) displacements incurred by particles within amorphous structures at mechanical equilibrium ($T=0$)~\cite{lutsko1989generalized,lemaitre2006sum}. At each coarse-grained box $m$, the local elastic modulus tensor ($X_{\alpha \beta \gamma \delta}$) is computed by subtracting the non-affine relaxation contribution ($X_{\alpha \beta \gamma \delta}^{NA\text{ }m}$) from the affine contribution ($X_{\alpha \beta \gamma \delta}^{A\text{ }m}$), expressed as:
\begin{equation}
X_{\alpha \beta \gamma \delta}^m
=X_{\alpha \beta \gamma \delta}^{A\text{ }m}-X_{\alpha \beta \gamma \delta}^{NA\text{ }m}.
\end{equation}

The contribution arising from affine deformation is measured through the ``Born'' term $X^{Bm}_{\alpha \beta \gamma \delta}$~\cite{born1955dynamical,alexander1998amorphous,lemaitre2006sum}, 
\begin{footnotesize}
\begin{equation}
    X_{\alpha \beta \gamma \delta}^{Bm}=\frac{1}{l_w^2} \sum_{i<j}\left(\frac{\partial^2 v}{\partial r_{i j}{ }^2}-\frac{1}{r_{i j}} \frac{\partial v}{\partial r_{i j}}\right) \frac{r_{i j \alpha} r_{i j \beta} r_{i j \gamma} r_{i j \delta}}{r_{i j}^2} \frac{q_{i j}}{r_{i j}}.
\end{equation}
\end{footnotesize}

To account for the non-zero values of the stress components in the random reference states of amorphous solids, a correction term ($X_{\alpha \beta \gamma \delta}^{Cm}$) is added to the contribution from affine deformations~\cite{barron1965second}:
\begin{footnotesize}
\begin{equation}
X_{\alpha \beta \gamma \delta}^{Cm} = -\frac{1}{2}\left(2 \sigma_{\alpha \beta}^{m} \delta_{\gamma \delta} - \sigma_{\alpha \gamma}^m \delta_{\beta \delta} - \sigma_{\alpha \delta}^m \delta_{\beta \gamma} - \sigma_{\beta \gamma}^m \delta_{\alpha \delta} - \sigma_{\beta \delta}^m \delta_{\alpha \gamma}\right).
\end{equation}
\end{footnotesize}
Therefore, the overall contribution of affine deformations is given by,
\begin{footnotesize}
\begin{equation}
X_{\alpha \beta \gamma \delta}^{Am}= X_{\alpha \beta \gamma \delta}^{Bm}+X_{\alpha \beta \gamma \delta}^{Cm}.
\end{equation}
\end{footnotesize}

\begin{figure}[]
\centering
  \includegraphics[width=1\columnwidth]{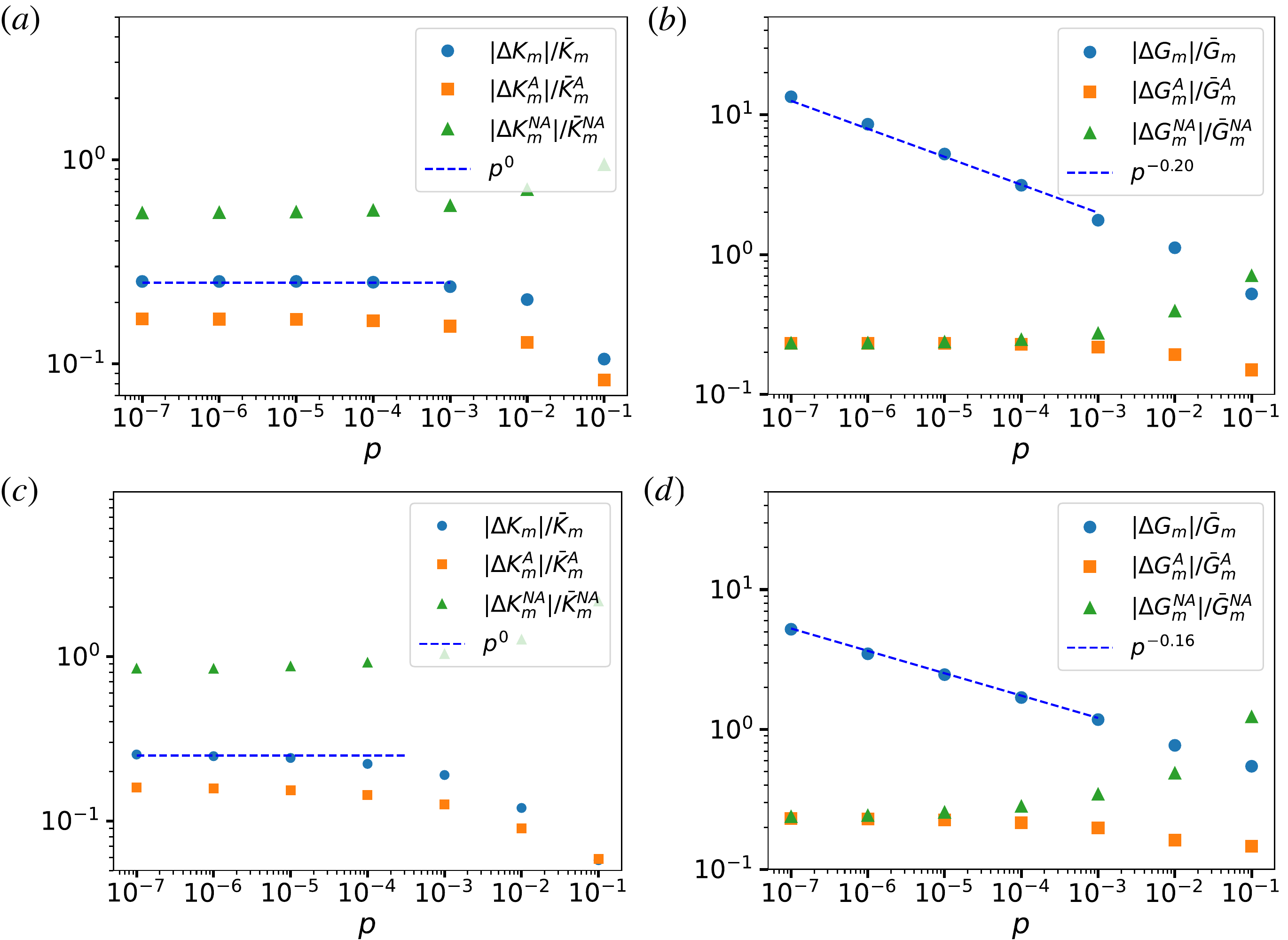}
  \caption{Relative fluctuations in the spatial distribution of local moduli as the unjamming transition is approached. For Harmonic interactions ($\alpha=2$), the relative fluctuations in local bulk modulus ($ \lvert \Delta K_m \rvert /\bar{K}_m$) remain constant, while relative fluctuations in local shear modulus ($\lvert \Delta G_m \rvert /\bar{G}_m$) grows as $p^{-0.20}$ at low pressures, as shown in \textbf{(a)} and \textbf{(b)} respectively. In the case of Hertzian interactions ($\alpha=5/2$), $\delta K_m /\bar{K}_m$ remains constant, and $\delta G_m /\bar{G}_m$ increases as $p^{-0.16}$, as shown in \textbf{(c)} and \textbf{(d)} respectively. The corresponding behaviors for the affine and nonaffine responses are represented by open triangles and squares, respectively. The local moduli are evaluated using a coarse graining length of $l_w = 3$.
}\label{relative_fluc_scaling}
\end{figure}
The application of strain to amorphous solids results in non-affine relaxation of particles, contributing to the overall modulus at mechanical equilibrium which is zero for the case of crystalline solids~\cite{lutsko1989generalized,lemaitre2006sum,leonforte2005continuum}. The non-affine contribution ($X_{\alpha \beta \gamma \delta}^{NAm}$) is determined through the analysis of the ``Hessian matrix'' in the Harmonic approximation of the potential energy at $T=0$~\cite{lemaitre2006sum,hentschel2011athermal}:
\begin{footnotesize}
\begin{equation}
\begin{aligned}
X_{\alpha \beta \gamma \delta}^{NAm} = \sum_{k} \frac{L^2}{\omega^{k^2}} \left(\sum_{i=1}^N \psi_{i}^{k} \cdot \frac{\partial \sigma_{\alpha \beta}^{m}}{\partial r_i}\right) \left(\sum_{j=1}^N \psi_{j}^{k} \cdot \frac{\partial \sigma_{\gamma \delta}}{\partial r_j}\right),
\end{aligned}
\end{equation}
\end{footnotesize}
where $\psi_{i}^{k}$ represents the $d$ dimensional displacement vector of particle $i$ in the normal mode $\psi^k$ of the ``Hessian matrix'' and $\omega_k^2$ is the corresponding eigenvalue.
The local bulk modulus of a two-dimensional solid within a coarse-grained box $m$ is measured as
\begin{equation}
    K_m= (X_{xxxx}^{m}+X_{yyyy}^{m}+X_{xxyy}^{m}+X_{yyxx}^{m})/4,
\end{equation}
and the local (simple) shear modulus at the $m$-th box is 
\begin{equation}
 G_m=  X^{m}_{xyxy}.
\end{equation}

Fig.~\ref{elastic_modulus_mean} illustrates the behavior of the mean local modulus as the pressure of the solids is decreased towards the unjamming point. The mean values of the local moduli remain largely independent of the coarse graining box sizes, here we employ coarse-grained boxes of linear length $l_w=3$. For Harmonic repulsion ($\alpha=2$) the mean bulk modulus ($\bar{K}_m$) remains constant, while the mean shear modulus ($\bar{G}_m$) scales as $p^{1/2}$ at low pressures, as shown in \textbf{(a)} and \textbf{(b)} respectively. In the case of Hertzian ($\alpha=5/2$) interactions, $\bar{K}_m$ scales as $p^{1/3}$ and $\bar{G}_m$ scales as $p^{2/3}$ as illustrated in \textbf{(c)} and \textbf{(d)} respectively. Notably, this scaling behaviour of the mean moduli with pressure is reminiscent  of scaling behavior observed in macroscopic moduli measured in model solids. Specifically, the bulk modulus scales as $p^{(\alpha-2)/(\alpha -1)}$ and the shear modulus scales as $p^{(\alpha-\frac{3}{2})/(\alpha -1)}$~\cite{o2003jamming,ellenbroek2006critical,vitelli2010heat}.

The behavior of the affine and non-affine components of these moduli reveals that the bulk modulus is mainly governed by its affine component, with a relatively small non-affine contribution. This explains the observed bulk modulus scaling, as the resistance to affine deformation scales with the double derivative of the potential energy ($\delta^{\alpha-2}$), with $\delta$ denoting the interparticle overlap. Consequently, as bulk deformations are primarily governed by an affine response, the pressure scales according to the force that scales as the derivative of the interaction energy ($\delta^{\alpha-1}$). On the other hand, the shear modulus exhibits a significant non-affine component, which is comparable to its affine counterpart, resulting in deviations from the expected scaling behavior~\cite{o2003jamming,ellenbroek2006critical,vitelli2010heat}. These findings regarding the interplay between the affine and non-affine response align with earlier studies~\cite{schlegel2016local,mizuno2016spatial}.

\begin{figure*}
 \centering
 \includegraphics[width=2\columnwidth]{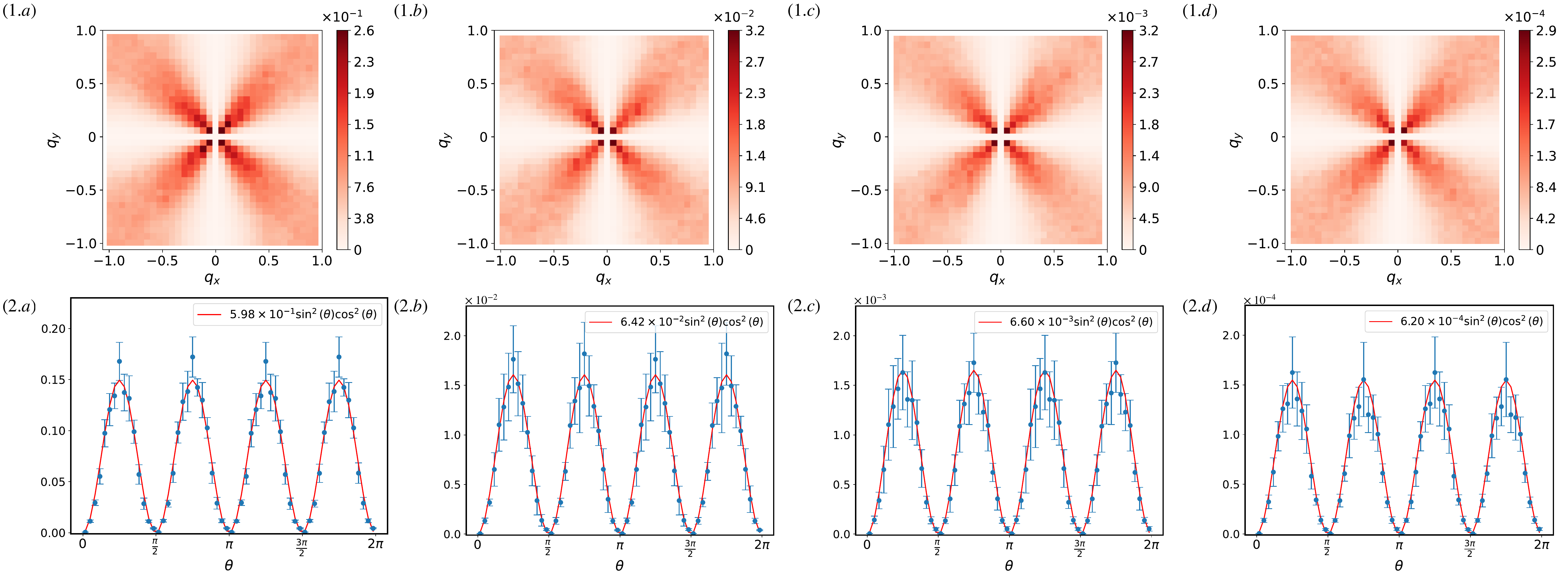}
\caption{Fourier space measurements of shear modulus correlations ($\langle \Delta G_{\vec{q}} \Delta G_{-\vec{q}}\rangle$) within two-dimensional amorphous packings with Hertzian repulsion at different pressures: (\textbf{a}) $p=10^{-3}$, (\textbf{b}) $p=10^{-4}$, (\textbf{c}) $p=10^{-5}$, and (\textbf{d}) $p=10^{-6}$. The shear modulus in real space is determined using coarse-grained boxes of size $l_w = 3$. The top row presents a two-dimensional representation of the shear modulus correlations. Notably, pinch point singularities are observed in the shear modulus correlations as $\lvert q \rvert \rightarrow 0$ for all pressures, indicating the presence of long-range anisotropic correlations. The bottom row displays the angular dependence of the correlations at small $\langle q \rangle$, well-fitted by the functional form $C_G(q, \theta) = g \sin^2\theta \cos^2\theta$ for different $g$ values at varying pressures. Angular dependency is computed through radial averaging in the range $\lvert q \rvert = \frac{2 \pi}{L}$ to $10\frac{2 \pi}{L}$ with angular binning intervals of $\frac{2\pi}{60}$. }
 \label{corr_tot_simple_shear}
\end{figure*}

In Fig.~\ref{relative_fluc_scaling}, we present the relative fluctuations in local moduli distributions as the unjamming transition is approached. The relative fluctuations in the bulk modulus remain constant, i.e., $\langle \lvert \Delta K_m \rvert \rangle \sim \langle \bar{K}_m \rangle$, as pressure $p \rightarrow 0$ for both Harmonic ($\alpha=2$) and Hertzian ($\alpha=2.5$) interactions, as shown in \textbf{(a)} and \textbf{(c)}, respectively. Conversely, the relative fluctuations in the local shear modulus ($\langle \lvert \Delta G_m \rvert \rangle / \langle \bar{G}_m \rangle$) increase near the unjamming point ($p\rightarrow0)$. For Harmonic interactions, $\langle \lvert \Delta G_m \rvert \rangle / \langle \bar{G}_m \rangle$ scales as $p^{-0.20}$ at low pressures, while for Hertzian interactions, it scales as $p^{-0.16}$, as illustrated in \textbf{(b)} and \textbf{(d)}. Here, $\langle \lvert \Delta K_m \rvert \rangle$ and $\langle \lvert \Delta G_m \rvert \rangle$ represent the standard deviation of the spatial fluctuations in the bulk and shear modulus, respectively, and their spatial averages are denoted by $\bar{K}_m$ and $\bar{G}_m$. 
The bulk modulus is primarily dominated by its affine component, with the non-affine contribution remaining relatively small, leading to a consistent behavior in its fluctuations. On the other hand, the total shear modulus, characterized by significant non-affine components, exhibits anomalous behavior in its fluctuations.

Quasi-localized excitations are crucial for understanding the anomalous mechanical and transport properties of amorphous solids~\cite{manning2011vibrational, widmer2008irreversible, lerner2021low, mizuno2014acoustic, mizuno2018phonon}. Therefore, it is important to study local elastic responses at length scales similar to those of floppy modes. In the next section, we explore the correlation between these local elastic responses. The presence of quasi-localized excitations leads to non-affine relaxation of particles when deformed in mechanical equilibrium. At small length scales, the non-affine response becomes substantial, resulting in regions with negative modulus and large relative fluctuations in local shear moduli distribution. This effect diminishes at larger length scales. A crossover length scale can be identified, beyond which the relative fluctuations become negligible. Following Mizuno~\textit{et al.}~\cite{mizuno2016spatial}, we identify a length scale that diverges as the system approaches $p \rightarrow 0$, exhibiting similar scaling behavior as the isostatic length scale in two dimensions~\cite{lerner2014breakdown}, as detailed in the Supplemental Material.

\section{Pinch Point Singularities in Shear Modulus Correlations} \label{Shear_modulus_correlation}
Previous studies investigating the local moduli of three-dimensional jammed solids and low-temperature glasses, using the ``fully local'' approach, suggested negligible spatial correlations between elastic moduli at large distances~\cite{mizuno2016spatial,shakerpoor2020stability}. In contrast, studies focusing on the acoustic properties of low-temperature glasses proposed the presence of long-range spatial correlations in elastic properties, determining local elastic moduli through vibrational dynamics~\cite{gelin2016anomalous}. However, Mizuno~\textit{et al.}~\cite{mizuno2018phonon}, utilizing a protocol similar to that in~\cite{gelin2016anomalous} for measuring affine elastic moduli, reported the absence of long-range spatial correlations in elastic moduli. 
Our study is closely related to the recent work of Zhang~\textit{et al.} who investigated local elastic moduli in jammed solids using actual strain-based measurements and observed long-range behavior beyond a specific pressure~\cite{zhang2023local}. In contrast, our study is based on linear response theory~\cite{lutsko1989generalized,lemaitre2006sum}, avoiding actual strain-based measurements that could introduce non-linearities in the response, particularly in fragile solids approaching the unjamming transition. Using linear response measurements, we aim to explore the local moduli correlations by conducting measurements of the fluctuations in the local simple shear modulus ($G_m$) within two-dimensional jammed packings. To analyze behavior across larger length scales, we employ Fourier space measurements and evaluate correlations at small wavenumbers. We utilize the ``fully local'' approach, as previously employed by~\cite{mizuno2016spatial}, considering both affine and non-affine contributions to the modulus.

We measure the correlation in Fourier space by performing discrete Fourier transform of $\Delta G_m$, which represents the difference between $G_m$ and its spatially averaged value $\bar{G}_m$:
\begin{equation}
 \Delta G(\vec{q})= \sum_{m}\exp(i \vec{q}.\vec{r}_m ) \Delta G_m.
\end{equation}
The modulus correlation in Fourier space is given by,
\begin{equation}
C_G(\vec q)= \langle\Delta G(\vec q )\Delta G(-\vec q )\rangle,
\end{equation}
where `$\langle\rangle$' denotes average over configurations.

We investigate pressure-dependent shear modulus correlations ($\langle G_{\vec{q}} G_{-\vec{q}}\rangle$) within two-dimensional amorphous packings with hertzian repulsion, exploring correlations across a range of pressures up to the unjamming point. Fig.~\ref{corr_tot_simple_shear} illustrates the correlations in Fourier space corresponding to six distinct pressures: (\textbf{a}) $p=10^{-3}$, (\textbf{b}) $p=10^{-4}$, (\textbf{c}) $p=10^{-5}$, and (\textbf{d}) $p=10^{-6}$. To obtain the shear modulus in real space, we employ coarse graining boxes of linear length $l_w = 3$. Notably, across all pressures we observe pinch point singularities in the shear modulus correlations at $\lvert q \rvert \rightarrow 0$. This demonstrates that the local shear modulus exhibits long-range correlations that are anisotropic and persist down to the unjamming transition. Interestingly, the structure of shear modulus correlations is identical to shear stress correlations observed in jammed solids~\cite{nampoothiri2020emergent,nampoothiri2022tensor}. 

These correlations exhibit an angular dependence of 
\begin{equation}
    C_G(q, \theta) = g \sin^2\theta \cos^2\theta,\label{eq_aniso_moduli_corr}
\end{equation}
as $\lvert \vec{q} \rvert \rightarrow 0$, where $\theta$ represents the angle and $g$ signifies the amplitude of the anisotropic correlations. The bottom row of Fig.~\ref{corr_tot_simple_shear} presents the angular dependency of these correlations at small $\lvert q \rvert$. The angular dependency is obtained through radial averaging in the range of $\lvert q \rvert = 2\frac{2 \pi}{L}$ to $10\frac{2 \pi}{L}$ with angular binning intervals of $\frac{2\pi}{60}$ to reduce noise. 

\begin{figure}[t!]
\centering
\includegraphics[width=1\columnwidth]{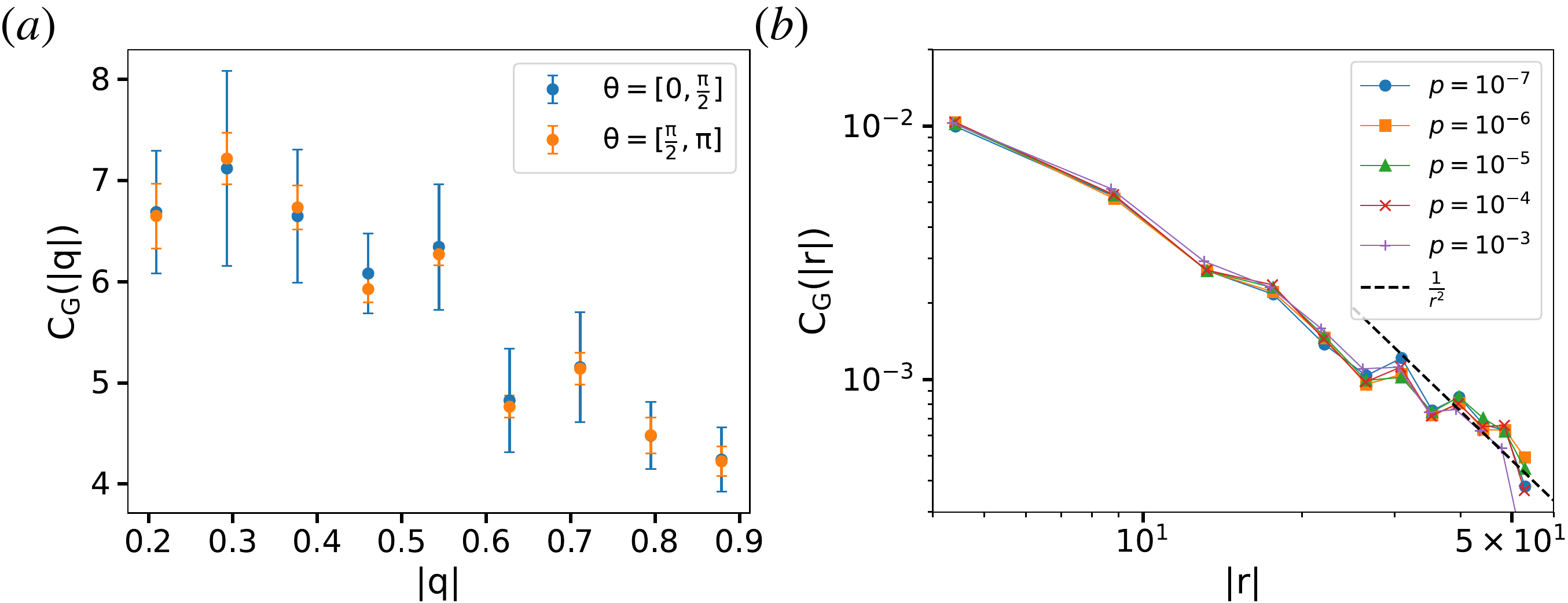}
\caption{\textbf{(a)} $C_G(|q|)$ as a function of $|q|$ for packings of particles interacting via Hertzian interaction at a pressure of $p = 10^{-6}$. The plot shows the integration of $C_G(|q|, \theta)$ over the angular ranges $\theta = \left[0, \frac{\pi}{2}\right]$ and $\theta = \left[\frac{\pi}{2}, \pi\right]$. As $|q| \rightarrow 0$, the correlation is independent of $|q|$ near the origin. \textbf{(b)} Integrated correlation in real space $C_r$ (defined in \eqref{eq_radial_corr}) as a function of $|r|$. The correlation decays as $\sim \frac{1}{r^2}$ at large distances.
}\label{fig_radial_corr_in_Fourier}
\end{figure}
\begin{figure}[t!]
\centering
\includegraphics[width=1\columnwidth]{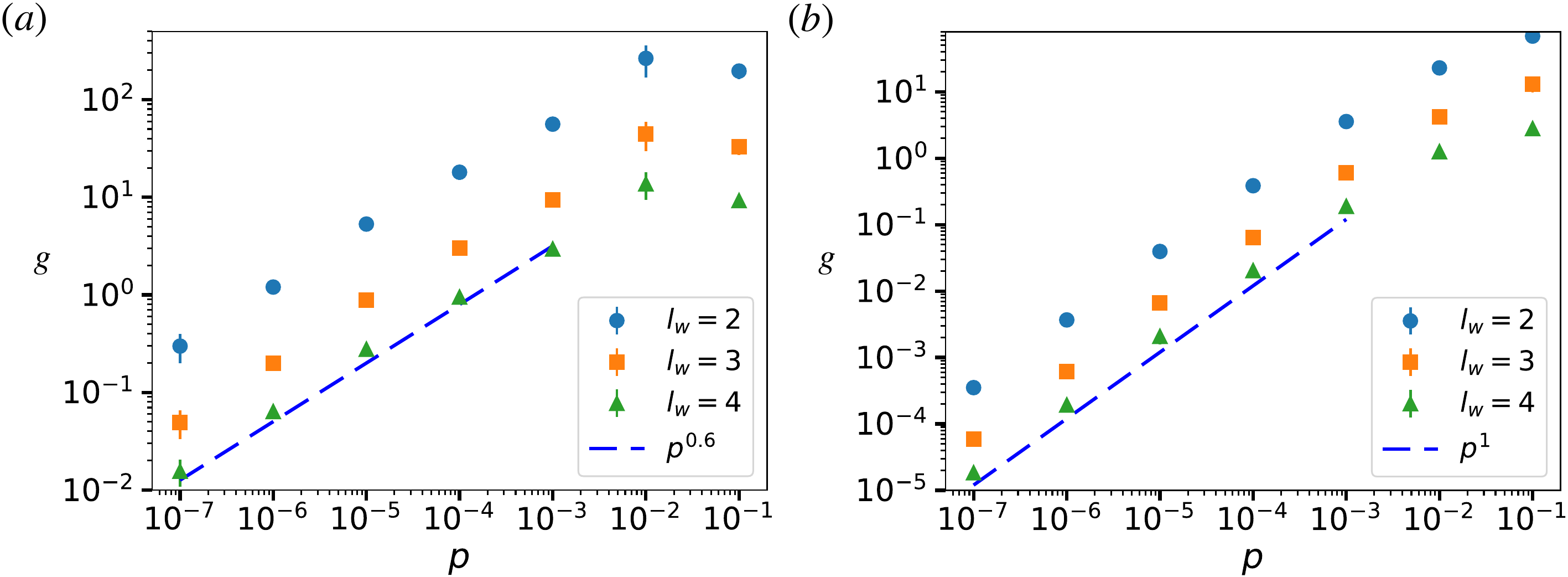}
  \caption{
  Dependence of the amplitude ($g$) of shear modulus correlations, which behave as $C_{G}(q,\theta) = g \sin^2\theta \cos^2\theta$, on the pressure ($p$) of the packings for different model solids, measured employing various coarse-graining lengthscales ($l_w$). \textbf{(a)} For $\alpha = 2$, the amplitude displays a decay proportional to $p^{0.6}$ as the pressure nears the unjamming point. \textbf{(b)} $\alpha = \frac{5}{2}$ exhibits a power-law behavior of $p^{1}$ near the unjamming point.
  }
\label{anIso_shear_modulus_correlations}
\end{figure}

To illustrate the independence of the Fourier space correlation $C_G(\lvert q \rvert, \theta)$ with respect to $\lvert q \rvert$, we measure $C_G(|q|)$ by integrating $C_G(|q|, \theta)$ over the angular range for different values of $|q|$, specifically from $|q|$ to $|q| + d|q|$, where $d|q| = \frac{2\pi}{L}\sqrt{2}$. This integration is performed separately for the first quadrant ($\theta = [0, \frac{\pi}{2}]$) and the second quadrant ($\theta = [\frac{\pi}{2}, \pi]$).

Fig.~\ref{fig_radial_corr_in_Fourier} \textbf{(a)} illustrates $C_G$ as a function of $|q|$ for packings of particles at a pressure of $p = 10^{-6}$. As $|q| \rightarrow 0$, the correlation becomes independent of $|q|$ and depends only on the angular coordinate. This indicates that in real space, the correlations decay as $1/r^2$ at large length scales. To illustrate this, we measure the integrated correlation function by performing the angular integration while taking into account the anisotropic nature of the correlation function,
\begin{equation}
C_G(|r|) = -\frac{1}{2\pi} \int_0^{2\pi} \frac{\langle G(\vec{r}) G(0) \rangle}{\langle G(0) G(0) \rangle} \cos(4\theta) d\theta .\label{eq_radial_corr}
\end{equation}

Fig.~\ref{fig_radial_corr_in_Fourier} \textbf{(b)} demonstrates that $C_G(|r|)$ decays as $\frac{1}{r^2}$ at all pressures ($p$) up to the unjamming point.

Next, we investigate the behavior of modulus correlations across different model solids by examining the correlation strength as the pressure decreases towards the unjamming point. Fig.~\ref{anIso_shear_modulus_correlations} illustrates the dependence of $g$, the amplitude of anisotropic shear stress correlation defined in \eqref{eq_aniso_moduli_corr}, on the pressure $p$ of the packings for various model solids, utilizing different coarse-graining box sizes for modulus measurement in real space. Notably, an increase in the coarse-graining box size leads to reduced fluctuations ($\Delta G_m$), thereby affecting the correlation magnitude, while the scaling behavior of the correlation remains consistent. In \textbf{(a)}, for $\alpha = 2$, we observe that the correlation decays as $p^{0.6}$ when the pressure decreases towards the unjamming point. Conversely, in \textbf{(b)} for $\alpha = \frac{5}{2}$, the correlation displays a power-law behavior of $p^{1}$. Further details regarding the shear modulus correlation strength across various $\alpha$ values can be found in the Supplemental Material. However, we do not discern a consistent relationship between the decay exponent of correlation strength as the system approaches the unjamming transition and $\alpha$ in the case of shear modulus.

\subsection{Correlation of the Affine Responses}\label{appen_affine_response}

\begin{figure}[]
\centering
\includegraphics[width=1\columnwidth]{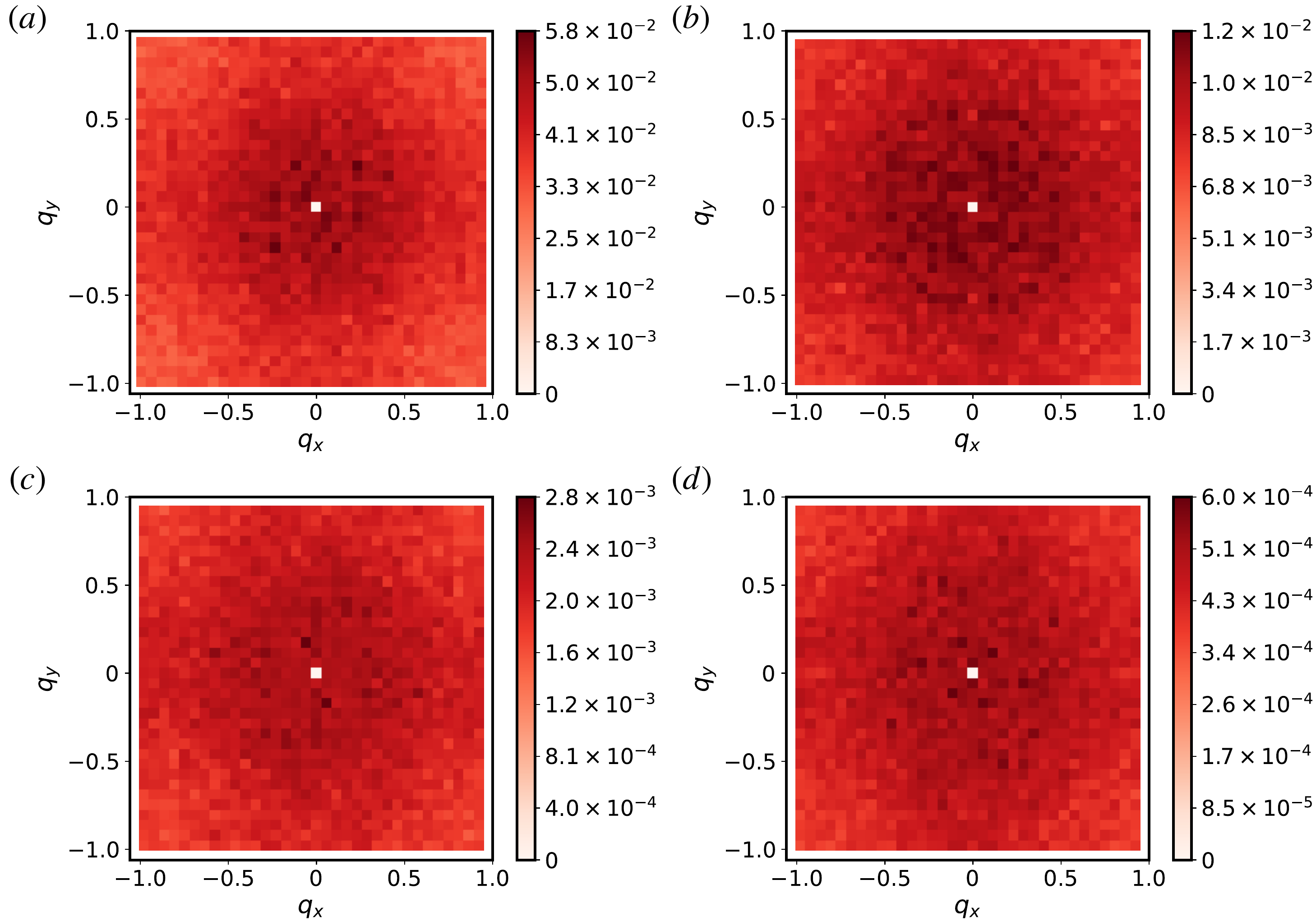}
\caption{Fourier space measurements of the correlation of affine shear modulus within two-dimensional amorphous packings of Hertzian repulsion under varying pressures: (\textbf{a}) $p=10^{-3}$, (\textbf{b}) $p=10^{-4}$, (\textbf{c}) $p=10^{-5}$, and (\textbf{d}) $p=10^{-6}$. The correlation remains independent of $\vec{q}$, exhibiting isotropic and radially independent characteristics. The modulus is measured using coarse-grained boxes of size $l_w = 3$. The correlations in Fourier space are independent of $\vec{q}$, revealing isotropic and radially independent characteristics, implying minimal correlation at significant spatial distances.}
\label{affine_shear_correlation_figure}
\end{figure}

Next, we investigate the correlation of the local affine shear modulus, measured in Fourier space by conducting a discrete Fourier transform of $\Delta G_{m}^A$, representing the difference between $G_{m}^A$ and its spatially averaged value $\bar{G}_{m}^A$,
\begin{equation}
 \Delta G_{m}^A(\vec{q})= \sum_{m}\exp(i \vec{q}.\vec{r}_m ) \Delta G_{m}^A.
\end{equation}
The correlation in Fourier space is given by,
 \begin{equation}
 C_{G}^A(\vec q ) = \langle\Delta G_{m}^A(\vec q )\Delta G_{m}^A(-\vec q )\rangle.
 \end{equation}

The correlation of the local affine shear modulus under various pressures is illustrated in Fig.~\ref{affine_shear_correlation_figure}. In Fourier space the affine modulus correlations appear independent of $\vec{q}$, displaying isotropic and radially independent behaviour near $\lvert q \rvert \rightarrow 0$. This implies a negligible correlation in the affine shear modulus across significant spatial distances, consistent with findings by Mizuno \textit{et al.}~\cite{mizuno2018phonon}, despite employing a different measurement protocol. Upon considering non-affine displacements of particles to maintain the mechanical equilibrium of solids, long-range spatial correlations emerge in the elastic shear modulus, resembling structures observed in stress correlations.

\section{Conclusion and Discussion}
\label{sec_conclusion}

In conclusion, our study reveals the presence of long-ranged anisotropic correlation in the local elastic response of amorphous solids in mechanical equilibrium. While previous investigations into local moduli often suggested negligible spatial correlations between elastic moduli at large distances~\cite{mizuno2016spatial, shakerpoor2020stability}. On the other hand, studies focusing on the acoustic properties of low-temperature glasses, such as by Gelin~\textit{et al.}~\cite{gelin2016anomalous} proposed the existence of long-range spatial correlations in elastic properties. However, it must be noted that this study measured the \textit{affine response}, and due to the specific form of the potential utilized in their study (inverse power-law), the affine response exhibited exact proportionality to stress, resulting in the observation of long-range correlations similar to stress correlations. Mizuno~\textit{et al.}~\cite{mizuno2018phonon} highlighted this issue and reported the absence of long-range correlation in affine moduli by measuring the affine response following the formalism used by Gelin~\textit{et al.}~\cite{gelin2016anomalous}. In a more recent study, Mahajan \textit{et al.} investigated the response of particle-level stresses to {\it global} strain deformations followed by energy minimization, and demonstrated that local moduli measured in this manner indeed exhibit long-ranged anisotropic correlations~\cite{mahajan2021emergence}.

In this study, we investigated long-range correlations in local elastic moduli within two-dimensional granular packings of soft repulsive disks. Employing the ``fully local'' approach to compute the local moduli~\cite{mizuno2013elastic, mizuno2016spatial}, we considered both affine and non-affine contributions to the modulus. By analyzing the fluctuations in the local simple shear modulus ($G_m$) in Fourier space, we unveil the existence of long-range correlations in the local elastic properties of such athermal materials. These correlations display anisotropic characteristics with pinch-point singularities in Fourier space, consistent with recent observations in real space by Zhang \textit{et al.}~\cite{zhang2023local}. Our study reveals that these correlations in Fourier space are similar to the correlations in the shear components of the local stress tensor and exhibit a power-law decay similar to the stress correlations, following a $\frac{1}{r^d}$ decay in $d$ dimensions, in real space.

Our investigation into the correlations within the affine response indicates negligible correlations at large spatial distances. However, upon considering the response arising from non-affine displacements of particles associated with vibrations about mechanical equilibrium in solids, long-range spatial correlations emerge in the elastic shear modulus, resembling the structures observed in stress correlations. The bulk modulus correlations are short-ranged, being independent of $\vec{q}$ as evidenced by Fourier space analysis, akin to pressure correlations observed in amorphous solids. These observations suggest an intriguing relationship within such solids: linking stiffness to deformations and tensions at mechanical equilibrium, which warrants further investigation.

Furthermore, our study quantitatively explores the behaviour of the correlation near the unjamming transition across model solids with short-range repulsive potentials. We showed that stress correlations exhibit a universal quadratic decay ($p^2$) as the pressure approaches the unjamming point, irrespective of the specific interaction details. In contrast, modulus correlations exhibit a model-specific behavior, displaying power-law decays with exponents related to the specific interaction potential. Fluctuations in the local affine moduli scale with their spatial average, i.e. $\langle |\Delta K_m| \rangle \sim \langle \bar{K}_m \rangle$, as pressure $p$ decreases towards the unjamming point. In contrast, the total shear modulus, which includes significant non-affine contributions, exhibits anomalous scaling behavior in its spatial fluctuations. The relative fluctuations in the local shear modulus ($\langle |\Delta G_m| \rangle / \langle \bar{G}_m \rangle$) increase according to distinct power-law exponents as the system approaches the unjamming transition. For Harmonic interactions ($\alpha = 2$), $\langle |\Delta G_m| \rangle / \langle \bar{G}_m \rangle$ scales as $p^{-0.20}$ at low pressures, whereas for Hertzian interactions ($\alpha = 2.5$), it scales as $p^{-0.16}$. While, the spatial average of the shear modulus scales as $\langle \bar{G}_m \rangle \sim p^{(\alpha - \frac{3}{2}) / (\alpha - 1)}$. These findings highlight the critical influence of non-affine deformations on the mechanical properties of amorphous solids.

Finally, our findings challenge the prevailing assumptions in theories that attribute the anomalous properties of amorphous solids to spatially heterogeneous moduli, presuming negligible correlations beyond small length scales. Although many previous studies have observed long-range stress correlations in low-temperature glasses~\cite{lemaitre2014structural,lemaitre2018stress,tong2020emergent}, some studies have attributed this behaviour to the percolation of a force-bearing network~\cite{tong2020emergent}. In this context, it will be interesting to study the correlation in stiffness within glasses as temperature decreases in order to ascertain whether the long-range correlations observed under mechanical equilibrium conditions coincide with the formation of a force-bearing network.
\\

\begin{acknowledgments}
We thank Bulbul Chakraborty, Vishnu V. Krishnan, Jishnu N. Nampoothiri and Roshan Maharana for useful discussions. The work of K.~R. was partially supported by the SERB-MATRICS grant MTR/2022/000966. This project was funded by intramural funds at TIFR Hyderabad under Project Identification No. RTI 4007 from the Department of Atomic Energy, Government of India.
\end{acknowledgments}

\bibliographystyle{apsrev4-2} 
\bibliography{heterogeneity}

\clearpage

\onecolumngrid

\setcounter{section}{0}  
\renewcommand{\thesection}{S\arabic{section}}  

\section*{Supplemental Material for\\ ``Long-Range Correlations in Elastic Moduli and Local Stresses at the Unjamming Transition''}

    In this Supplemental Material, we provide additional details related to the results presented in the main text. We provide measures of the bulk modulus correlations in Fourier space at different pressures. We analyze the behavior of the elasticity lengthscale at the approach to unjamming. We provide measures of the correlations of the shear modulus in real space. Finally, we analyze the behavior of the unjamming transition for systems with different repulsive interactions.

\section{Bulk Modulus Correlation}\label{appen_bulk_response}

\begin{figure}[h]
\includegraphics[width=0.75\columnwidth]{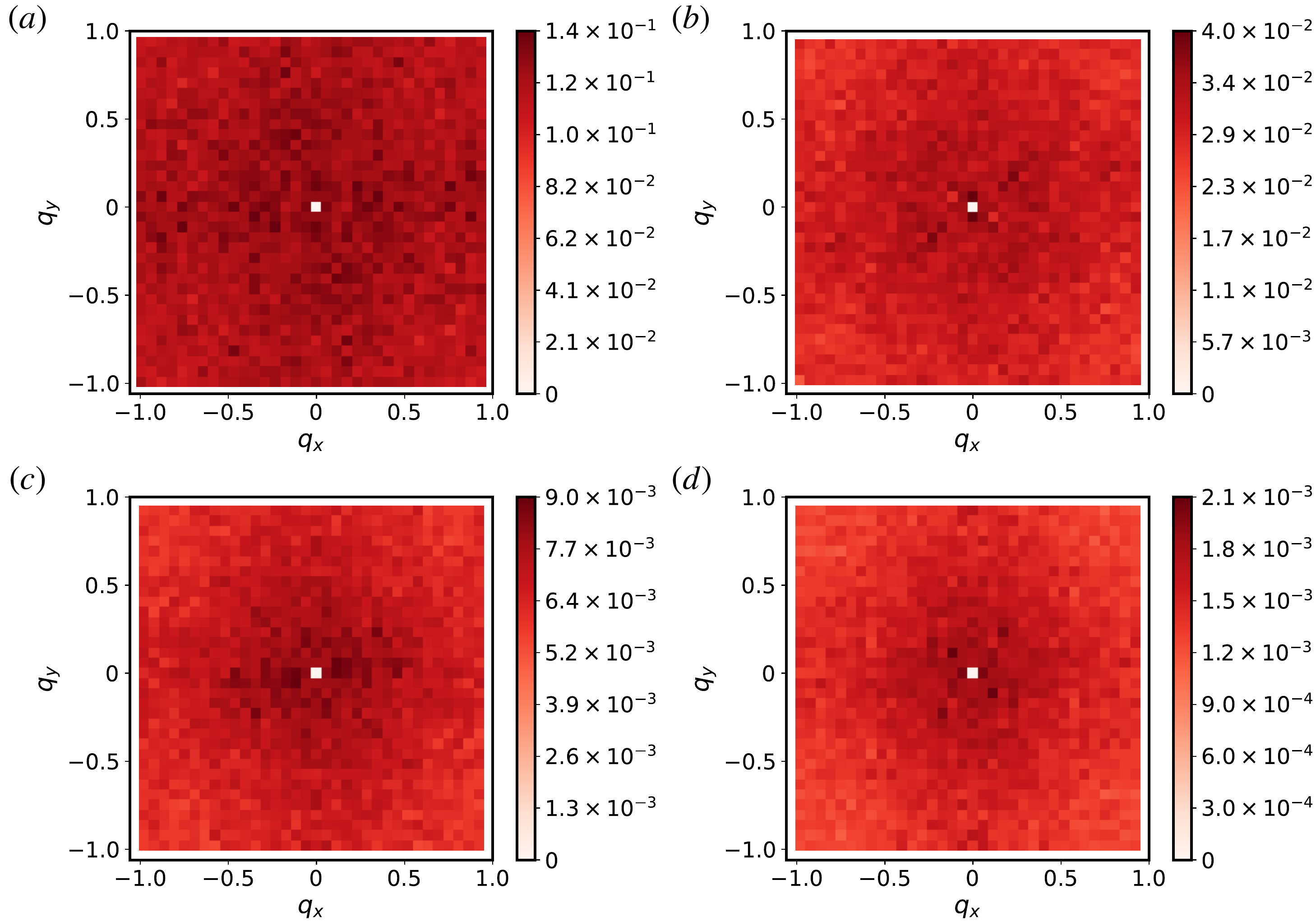}
\caption{Bulk modulus correlations in Fourier space for two-dimensional amorphous packings of $N=8192$ disks with Hertzian repulsion at various pressures: (\textbf{a}) $p=10^{-3}$, (\textbf{b}) $p=10^{-4}$, (\textbf{c}) $p=10^{-5}$, and (\textbf{d}) $p=10^{-6}$. The correlations are independent of $\vec{q}$, indicating negligible correlation at large distances. The local bulk modulus in real space is measured using coarse-grained boxes of size $l_w = 3$. }
\label{bulk_correlation_figure}
\end{figure}
In this section, we illustrate the bulk modulus correlation measured in Fourier space. The bulk modulus is primarily dominated by its affine component, with the non-affine contribution being relatively small. We perform a discrete Fourier transform of $\Delta K_{m}^A$, representing the difference between $K_{m}^A$ and its spatially averaged value $\bar{K}_{m}^A$,
\begin{equation}
\Delta K_{m}^A(\vec{q}) = \sum_{m} \exp(i \vec{q} \cdot \vec{r}_m ) \Delta K_{m}^A.
\end{equation}

The correlation in Fourier space is given by,
\begin{equation}
C_{G}^A(\vec{q}) = \langle \Delta G_{m}^A(\vec{q}) \Delta G_{m}^A(-\vec{q}) \rangle.
\end{equation}
Fig.~\ref{bulk_correlation_figure} illustrates the Fourier space representation of bulk modulus correlation across different pressures. In Fourier space, the correlations appear independent of $\vec{q}$, displaying isotropic and radially independent behavior near $\lvert q \rvert \rightarrow 0$. This implies a negligible correlation in local bulk modulus across significant spatial distances, similar to pressure correlations observed in amorphous solids~\cite{nampoothiri2022tensor}.

\section{Elasticity lengthscale of jammed solids}
The presence of quasi-localized excitations leads to non-affine relaxation of particles when deformed in mechanical equilibrium. At small length scales, the non-affine response becomes substantial, resulting in regions with a negative modulus.

\begin{figure}[]
\includegraphics[width=1\columnwidth]{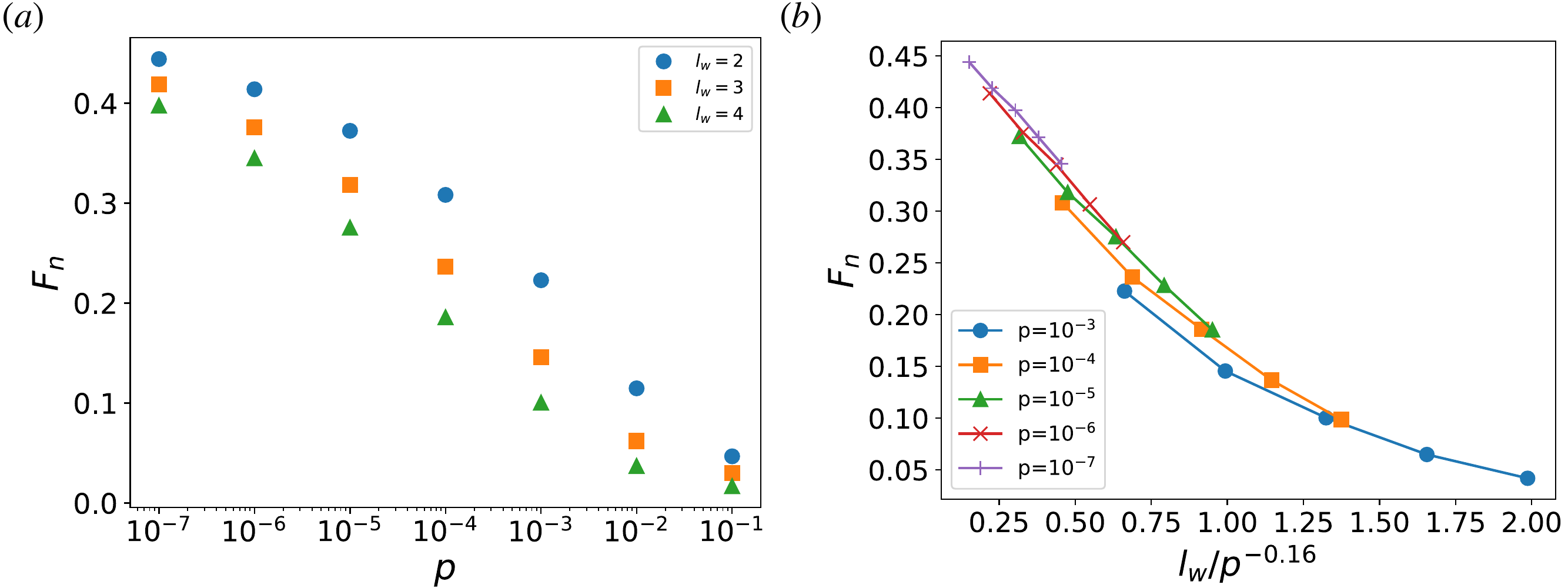}
\caption{(a) Fraction of negative shear modulus region as a function of pressure. (b) Scaling collapse of $F_n(l_w, p)$ reveals a length scale that diverges as $p^{-0.16}$ as the system approaches unjamming.}
\label{fig_neg_shear_modulus}
\end{figure}

In Fig.~\ref{fig_neg_shear_modulus}~\textbf{(a)}, we measure the fraction of regions with a negative shear modulus, defined as:

\begin{equation}
F_n = \int_{G_m < 0} P(G_m) , dG_m.
\end{equation}

As the unjamming transition is approached ($p \to 0$), $F_n$ increases, indicating a large number of unstable regions as the system becomes more fragile. The number of unstable regions decreases when measured using larger coarse-grained boxes, suggesting the existence of a length scale above which elasticity is recovered. A crossover length scale can be constructed beyond which this phenomenon occurs. From the scaling collapse of $F_n(p, l_w)$ shown in Fig.~\ref{fig_neg_shear_modulus}~\textbf{(b)}, we identify a length scale that diverges as $p^{-0.16}$ as the system approaches $p \to 0$, consistent with the isostatic length scale scaling in two dimensions~\cite{lerner2014breakdown}. Mizuno~\textit{et al.}~\cite{mizuno2016spatial} have reported the same scaling behavior for the length scale at which the relative fluctuations become negligible.


\begin{figure}[]
\includegraphics[width=0.75\columnwidth]{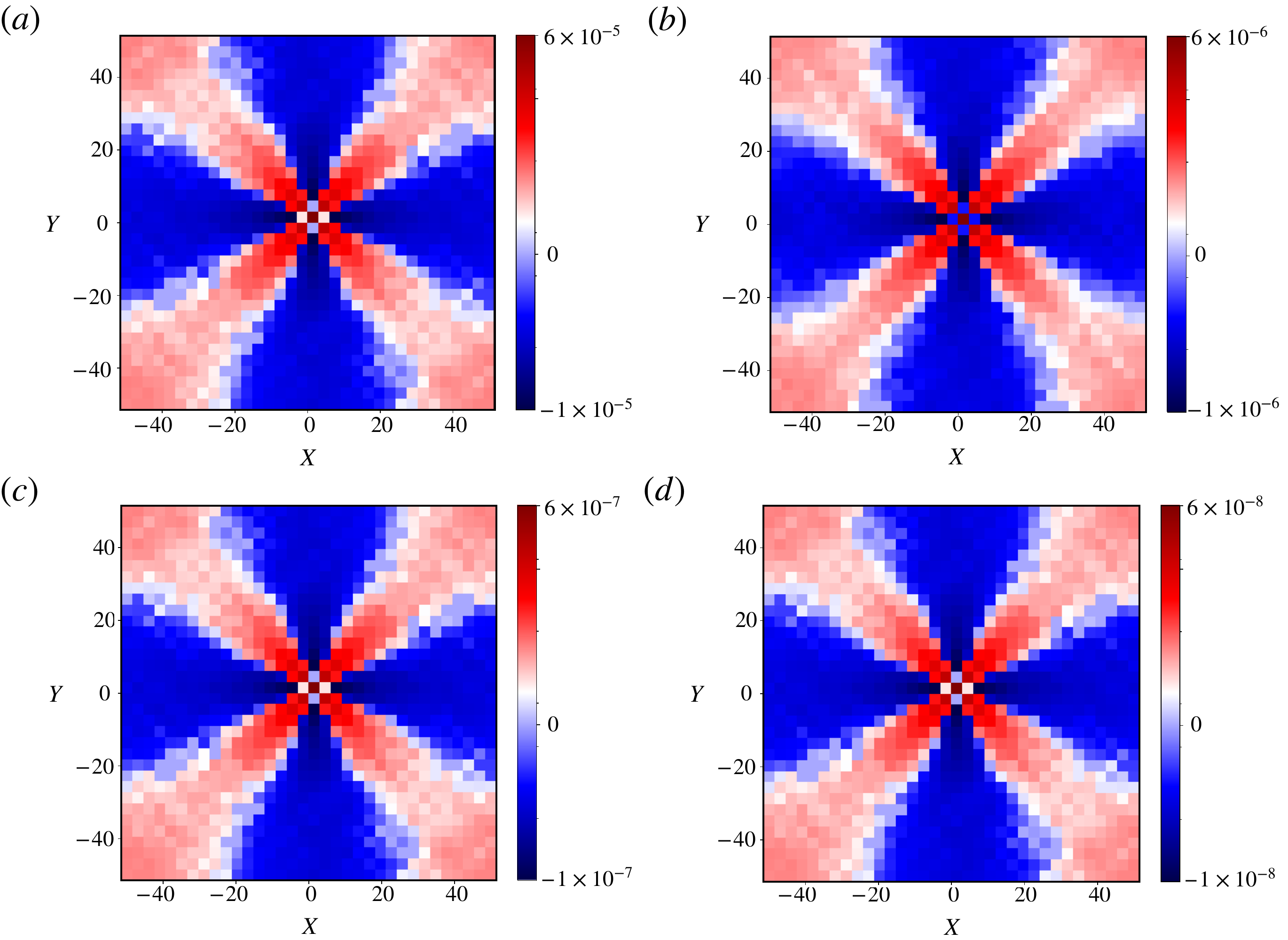}
\caption{The shear modulus correlation in real space for a packing of $N=8192$ disks with Hertzian repulsion at different pressures: \textbf{(a)} $p=10^{-3}$, \textbf{(b)} $p=10^{-4}$, \textbf{(c)} $p=10^{-5}$, and \textbf{(d)} $p=10^{-6}$. The local moduli are measured using coarse-grained boxes with a linear width of $l_w=3$. The local shear moduli exhibit long-range anisotropic correlations, manifesting a quadrupole shape at all pressures up to the unjamming point.}\label{corr_real_space}
\end{figure}

\section{Shear modulus Correlation in real space}
In this section, we present the shear modulus correlation in real space, $\langle \Delta G(\vec{r}) \Delta G(0) \rangle$. The shear modulus correlation exhibits long-range correlations with a quadrupole shape. This large-scale behavior is evident from the presence of non-vanishing anisotropic correlations as $|q| \rightarrow 0$. Fig.~\ref{corr_real_space} illustrates the real space correlations across different pressures for packings of $N=8192$ disks interacting via Hertzian interactions.

\section{The Unjamming Transition Behavior Across Various Repulsive Strength }\label{appen_scaling_for_other_alpha}

\begin{figure}[]
\includegraphics[width=0.75\columnwidth]{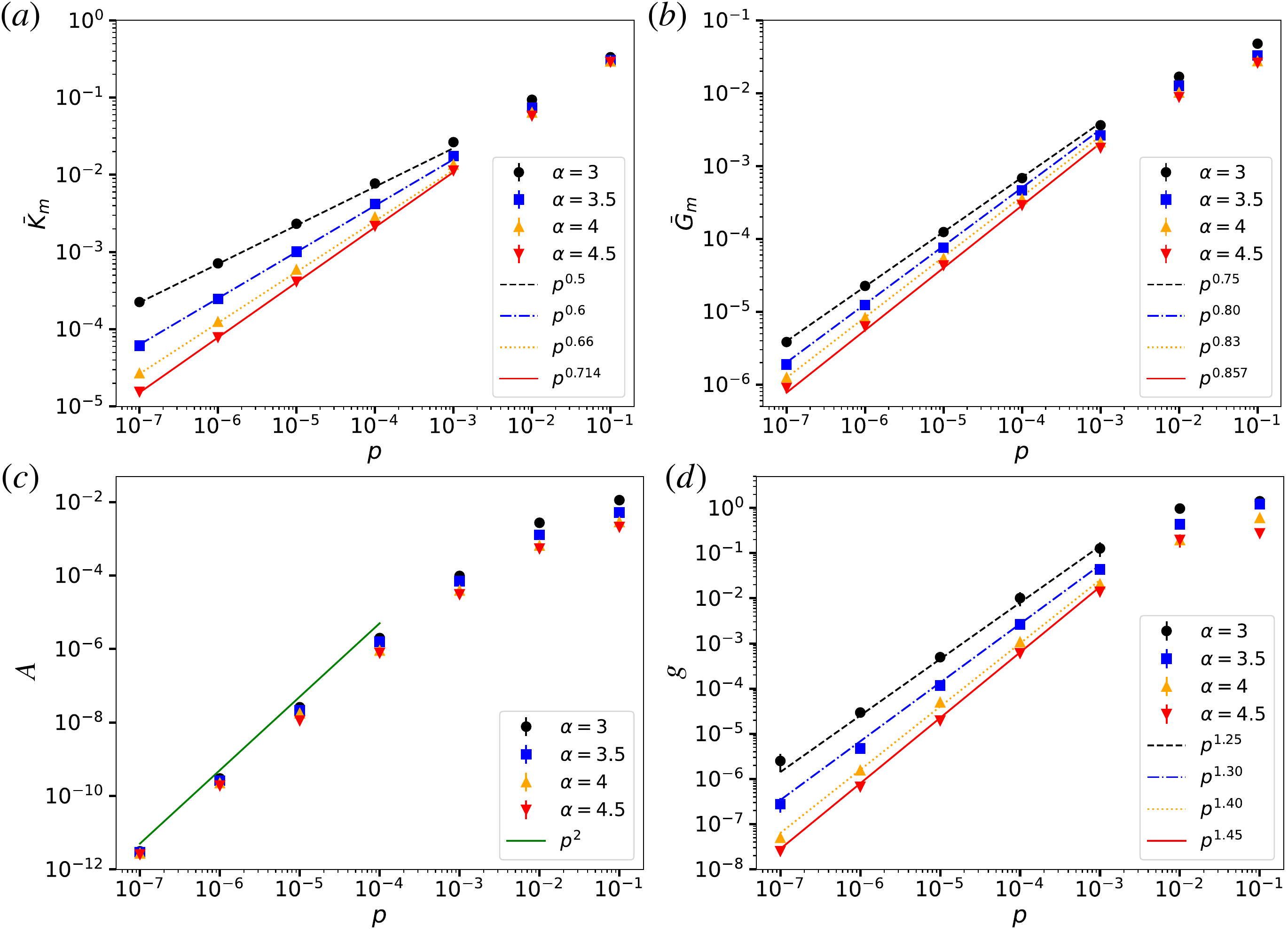}
\caption{Dependence of various characteristics on the pressure of the solids is illustrated for different values of $\alpha$: \textbf{(a)} mean bulk modulus ($\bar{K}_m$), \textbf{(b)} mean shear modulus ($\bar{G}_m$), \textbf{(c)} stress correlation strength ($A$), and \textbf{(d)} shear modulus correlation strength ($g$).}
\label{scalings_at_alpha_figure}
\end{figure}

\begin{table}
  \caption{Scaling behavior of mean bulk modulus ($\bar{K}_m$), mean shear modulus ($\bar{G}_m$), stress correlation strength ($A$), and shear modulus correlation strength ($g$) with pressure as the system approaches the unjamming transition across various repulsive strengths ($\alpha$). }
  \begin{ruledtabular}
   \begin{tabular}{ccccc}
         $\alpha$ & $\bar{K}_m$ & $\bar{G}_m$ & $A$ & $g$\\
         \hline
         $2$ & $p^{0}$ & $p^{0.5}$ & $p^{2}$ & $p^{0.5}$\\ 
         $2.5$ & $p^{0.33}$ & $p^{0.66}$ & $p^{2}$ & $p^{1}$\\
         $3$ & $p^{0.5}$ & $p^{0.75}$ & $p^{2}$ & $p^{1.25}$\\
         $3.5$ & $p^{0.6}$ & $p^{0.8}$ & $p^{2}$ &  $p^{1.30}$\\
         $4$ & $p^{0.66}$ & $p^{0.83}$ & $p^{2}$ & $p^{1.40}$\\
         $4.5$ & $p^{0.714}$ & $p^{0.857}$ & $p^{2}$  & $p^{1.45}$\\
    \end{tabular}
  \end{ruledtabular}
  \label{tab_scaling}
\end{table}


This section illustrates the unjamming transition behavior for various repulsive strengths ($\alpha$). Fig.~\ref{scalings_at_alpha_figure} presents the scaling behavior of different quantities as pressure approaches the unjamming point for distinct values of $\alpha$. Panels \textbf{(a)} through \textbf{(d)} in the figure show the scaling behaviour at the unjamming point for: \textbf{(a)} mean bulk modulus ($\bar{K}_m$), \textbf{(b)} mean shear modulus ($\bar{G}_m$), \textbf{(c)} stress correlation amplitude ($A$), and \textbf{(d)} shear modulus correlation amplitude ($g$). The scaling exponents associated with these quantities are listed in TABLE~\ref{tab_scaling}.

For all $\alpha$ values, the mean bulk modulus scales as $p^{(\alpha-2)/(\alpha-1)}$, while the mean shear modulus follows a scaling of $p^{(\alpha-\frac{3}{2})/(\alpha-1)}$, as shown in Fig.~\ref{scalings_at_alpha_figure} \textbf{(a)} and \textbf{(b)}, respectively. These power-law behaviors in the mean moduli are similar to those observed in the jamming of soft repulsive spheres~\cite{o2003jamming,ellenbroek2006critical,vitelli2010heat}.

Stress correlations behave as $C_{xyxy}(q,\theta) = A \sin^2\theta \cos^2\theta$ as $|q| \to 0$, with $A$ varying with the pressure ($p$) of the packings. Notably, stress correlations in jammed solids exhibit a universal characteristic, showing a quadratic decay of $A$, i.e. $A \sim p^2$ as pressure approaches the unjamming point, independent of specific interaction details, as shown in Fig.~\ref{scalings_at_alpha_figure}~\textbf{(c)}.
\begin{figure}[]
\includegraphics[width=1\columnwidth]{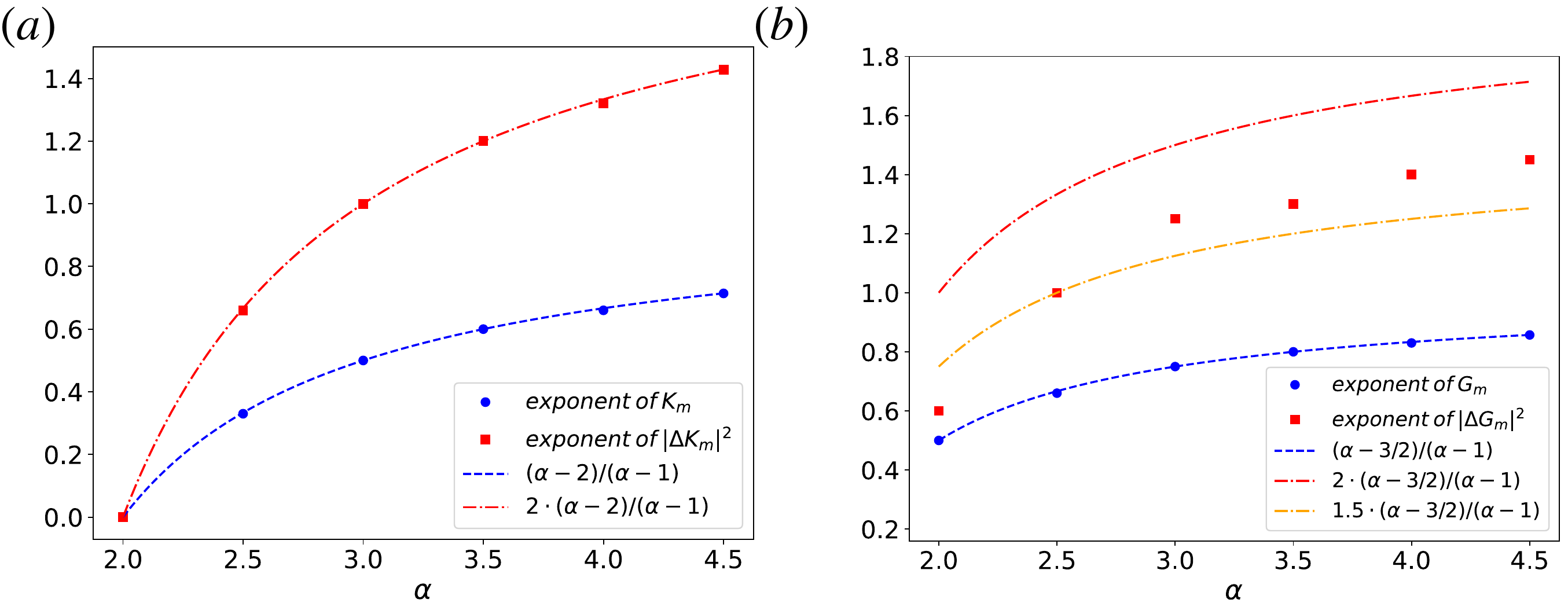}
\caption{\textbf{(a)} Bulk modulus fluctuations exhibit a scaling behavior of $p^{2(\alpha - 2)/(\alpha - 1)}$, while the mean bulk modulus scales as $p^{(\alpha - 2)/(\alpha - 1)}$. \textbf{(b)} In contrast, shear modulus fluctuations display anomalous behavior, lacking consistent correlation strength behavior with $\alpha$. However, the mean shear modulus scales as $p^{(\alpha - \frac{3}{2})/(\alpha - 1)}$.}
\label{scaling_vs_alpha_figure}
\end{figure}
Shear modulus correlation exhibits a similar structural pattern to shear stress correlation, $C_{G}(q,\theta) = g \sin^2\theta \cos^2\theta$ as $|q| \to 0$, with $g$ varying with the pressure ($p$) of the packings. However, the behavior of modulus correlations depends on the model, with $g$ decaying according to power-law exponents that vary based on the specific interaction potential, as illustrated in Fig.~\ref{scalings_at_alpha_figure}~\textbf{(d)}.

Fluctuations in the local bulk moduli scale with their spatial average; i.e. $|\Delta K_m|$ and $\bar{K}_m$ both scale as $p^{(\alpha-2)/(\alpha-1)}$ as pressure decreases towards the unjamming point for different values of $\alpha$, as shown in Fig.~\ref{scaling_vs_alpha_figure}~\textbf{(a)}. In contrast, the total shear modulus, which includes significant non-affine contributions, exhibits anomalous scaling behavior. Notably, no consistent relation is observed between the shear modulus fluctuations and the exponent of the interaction potential, as shown in Fig.~\ref{scaling_vs_alpha_figure}~\textbf{(b)}. However, the mean shear modulus scales as $p^{(\alpha-\frac{3}{2})/(\alpha-1)}$ as the unjamming transition is approached for different values of $\alpha$.



\end{document}